\documentclass[a4paper,11pt]{article}

\usepackage[utf8]{inputenc}
\usepackage[english]{babel}
\usepackage[myheadings]{fullpage}
\usepackage[T1]{fontenc}
\usepackage{subfig}
\usepackage{fancyhdr}
\usepackage{graphicx}
\usepackage{sectsty}
\usepackage{lipsum}
\usepackage{mathrsfs}
\usepackage{url}
\usepackage[dvipsnames, table, xcdraw]{xcolor}
\usepackage{amsmath}
\usepackage{amssymb}

\usepackage{bbold}
\usepackage{dsfont}
\usepackage{esint}
\usepackage{indentfirst}
\usepackage{xfrac}
\usepackage{lmodern}
\usepackage[T1]{fontenc}
\usepackage{lmodern}
\usepackage[T1]{fontenc}
\usepackage{slashed}
\usepackage{breqn}
\usepackage{autobreak}
\usepackage{setspace}
\usepackage[colorlinks]{hyperref}
\usepackage{microtype}
\usepackage{quoting}
\usepackage{float}
\usepackage{array}
\usepackage[export]{adjustbox}
\usepackage[section]{placeins}
\usepackage[capitalise]{cleveref}
\usepackage{enumitem}
\usepackage{scalefnt}
\usepackage{booktabs}
\usepackage{caption}
\usepackage{makecell}
\usepackage{authblk}

\usepackage[normalem]{ulem}

\usepackage{cite}

\usepackage{xspace}

\usepackage{colortbl}
\definecolor{myblue}{rgb}{0.8,0.85,1}
\definecolor{light-gray}{gray}{0.95}

\hypersetup{colorlinks=true,
citecolor=ForestGreen,
anchorcolor=CornflowerBlue,
linkcolor=NavyBlue,
urlcolor=Mahogany,
linkbordercolor={1 1 0}}

\usepackage{orcidlink}

\makeatletter
\newcommand{\thickhline}{\noalign {\ifnum 0=`}\fi \hrule height 1pt
\futurelet \reserved@a \@xhline}
\newcolumntype{"}{@{\hskip\tabcolsep\vrule width 1pt\hskip\tabcolsep}}
\makeatother

\allowdisplaybreaks

\def\beq {\begin{equation}}
\def\eeq {\end{equation}}
\def\bea {\begin{eqnarray}}
\def\eea {\end{eqnarray}}

\AtBeginDocument{}

\title{\LARGE {\bf \sffamily \boldmath Estimate of $\Gamma(H \to gg)$ at $\mathcal{O}(\alpha_s^7)$ from Pad\'e approximants with constraints from large-$n_f$}}

\author[a\,\protect\orcidlink{0000-0002-4426-7984}]{Diogo Boito}
\author[a\,\protect\orcidlink{0000-0001-5540-262X}]{Cristiane Y. London}
\author[a,b\,\protect\orcidlink{0000-0003-2864-5057}]{Guilherme A. Nogueira}

\affil[a]{\it Instituto de F\'isica de S\~ao Carlos, Universidade de S\~ao Paulo, CP 369, 13560-970, S\~ao Carlos, SP, Brazil}
\affil[b]{\it Instituto de F\'isica Gleb Wataghin, Universidade Estadual de Campinas, \newline R. S\'ergio Buarque de Holanda, 777, Campinas, São Paulo, Brazil\vspace{0.3cm}}

\date{}

\begin{document}
\begin{flushright}
{\small \today}
\end{flushright}

\vspace*{-0.7cm}
\begingroup
\let\newpage\relax
\maketitle
\endgroup
\date{}

\vspace*{-1.0cm}

\begin{abstract}
\noindent
In this work we estimate the first unknown QCD correction to the decay of the Higgs boson into gluons, $H\to gg$, in the heavy-top limit using Padé approximants built to the perturbative series and to its Borel transform. In the process, we also obtain estimates for the first unknown coefficients of the QCD $\beta$ function, of the QCD decoupling relation, and of the gluonium correlator. We also provide an updated prediction for the first unknown coefficient of $\Gamma(H\to b\bar{b})$. We determine the polynomial dependence of the perturbative coefficients on the number of quark flavors, $n_f$.
We point out that the large-$n_f$ result for $H\to gg$ can be obtained from results for the gluonium correlator already available in the literature.
The coefficient of the highest power of $n_f$ is then imposed as a constraint and is also used as a new test to discard unreliable approximants. The method is validated through the postdiction of the last known coefficient of the series. Our model-independent estimate for the yet unknown coefficient of order $\alpha_s^7$ of $\Gamma(H\to gg)$ is $-312\pm 15$ (in the $\overline{\mathrm{MS}}$ scheme with $\mu=m_H$, where $m_H$ is the Higgs mass). With our estimate the residual renormalization-scale dependence of $\Gamma(H\to gg)$ is reduced by a factor of almost 1.5.
\end{abstract}

\thispagestyle{empty}

\vspace*{0.0cm}

\setcounter{page}{1}

\vspace{1cm}
\newpage
\section{Introduction}

In the absence of direct evidence for new physics, precision measurements of Higgs decays at the LHC, the HL-LHC~\cite{ATLAS:2025eii}, and future machines, such as the FCC-ee~\cite{FCC:2025lpp,DelVecchio:2025gzw}, remain a compelling strategy for investigating physics beyond the Standard Model (SM). On the theory side, to meet the needs of future machines and increase the precision in the description of hadronic Higgs decays, there are two main challenges. First and foremost, the QCD parameters --- $\alpha_s$ and quark masses --- have to be determined with smaller uncertainties~\cite{Lepage:2014fla,Heinemeyer:2021rgq,Boito:2021scm}. Second, the decay widths and cross sections should be computed at higher orders in perturbative QCD (pQCD)~\cite{Heinemeyer:2021rgq}, especially if the projected reductions of the uncertainties in $\alpha_s(m_Z)$ and $m_H$ are achieved at a future FCC-ee~\cite{FCC:2025lpp}. With some observables already computed at five loops~\cite{Luthe:2016xec,Baikov:2016tgj,Herzog:2017ohr,Luthe:2017ttg}, computations at even higher orders are extremely demanding and are unlikely to become available in the near future. In this scenario, reliable estimates of higher-order perturbative coefficients~\cite{Ellis:1997sb,Boito:2018rwt,Boito:2021scm,Abbas:2022wnz} are welcome, since they can be used directly in phenomenology or to assess the error due to missing higher orders (MHOs) and the related residual renormalization-scale dependence.

The decays of the Higgs boson into hadrons constitute approximately $70\%$ of the SM decay width, and the predominant final states are bottom quarks and gluons~\cite{Djouadi:2005gi,Davies:2017rle}.
The decay $H\to gg$ (with radiation of additional partons implicit at higher orders), which is the second-largest hadronic channel and the main focus of this work, starts at order $\alpha_s^2$, and the pQCD corrections are organized in an expansion in $m_H^2/(4m_t^2)\approx 0.13$, where $m_H$ and $m_t$ are the Higgs and top-quark masses, respectively. The leading term in this expansion, which corresponds to taking $m_t\to\infty$ inside the loops, is known at next-to-next-to-next-to-next-to-leading order, N4LO, or $\mathcal{O}(\alpha_s^6)$~\cite{Chetyrkin:1997iv,Schreck:2007um,Baikov:2016tgj,Chetyrkin:2016uhw,Herzog:2017dtz}.
The decay $H\to b\bar{b}$ (also with implicit emission of additional partons) has the largest of all Higgs branching fractions. Its QCD corrections are known to N4LO, or $\alpha_s^4$ accuracy, in the $m_b/m_H\to 0$ limit --- a five-loop computation~\cite{Baikov:2005rw,Herzog:2017dtz}.

In this paper, we provide estimates for the first unknown perturbative coefficient of the decay of the SM Higgs boson into gluons, $H\to gg$, which is of $\mathcal{O}(\alpha_s^7)$, using Pad\'e approximants. In the process, we obtain estimates for the first unknown coefficients of the QCD $\beta$ function, of the QCD decoupling relation, of the imaginary part of the gluonium correlator, and of the Wilson coefficient of the effective Higgs--gluon coupling, obtained after integrating out the top quark, that governs $H\to gg$ as well as $gg\to H$. As a secondary application, we also update the estimate of the first unknown coefficient of $H\to b\bar b$.

To estimate the N5LO corrections we employ Padé approximants (PAs), also known as rational approximants, and some of their variants. The main advantage of PAs is that they provide a model-independent method to reconstruct higher orders in an expansion and, in some particular cases, the method is well rooted in mathematical theorems that guarantee convergence~\cite{POMMERENKE1973775}. In applications to pQCD series, one must depart from the canonical cases backed by theorems, but a body of literature supports the notion that these approximants are still successful in providing a guide to higher orders~\cite{Ellis:1995jv,Ellis:1996zn,Ellis:1997sb,Boito:2018rwt,Boito:2022fmn}.

Since the number of terms known exactly in pQCD series is small, it is important to validate the method. The main ideas were already presented in two other papers, where we provided estimates for unknown pQCD coefficients for $H\to b\bar b$~\cite{Boito:2021scm} and for $\tau\to {\rm hadrons}+\nu_\tau$~\cite{Boito:2018rwt}. In these papers, the large-$\beta_0$ limit of QCD, which provides a realistic toy model for the series to all orders in $\alpha_s$, was used to validate the strategy. Here, the validation is provided by the postdiction of the last known coefficient of each pQCD series. As in Refs.~\cite{Boito:2021scm,Boito:2018rwt}, we use more than one type of approximant: the usual PAs as well as D-log Padé approximants~\cite{Baker1996pade,Boito:2018rwt}, built both to the power series in $\alpha_s$ and to its Borel transform, a method known as Pad\'e-Borel~\cite{Baker:1976ff}.

Here, however, results obtained in the large-$\beta_0$ limit are not used to guide the approximation, and only exact large-$n_f$ information is employed. Large-$n_f$ results provide the coefficient of the highest power of $n_f$ at each order in $\alpha_s$ exactly, where $n_f$ is the number of light-quark flavors. These results have been known for a long time for $H\to b\bar b$~\cite{Broadhurst:2000yc,Jamin:2016ihy}. Here we show that the large-$n_f$ results for $H\to gg$ can also be extracted from results already existing in the literature~\cite{Jamin:2012nn}, since the highest power of $n_f$ in its perturbative coefficients stems entirely from the gluonium correlator. We use this information as input for the prediction of unknown coefficients as polynomials in $n_f$, as done in some older works~\cite{Ellis:1996zn,Ellis:1997sb}, and also as a new test that is used to discard approximants whose large-$n_f$ behavior is incompatible with the exact result.
In the case of $H\to gg$, instead of building approximants directly to the decay rate, we apply the method separately to the imaginary part of the gluonium correlator, to the decoupling relation, and to the $\beta$ function, and combine the results through the low-energy theorem and the renormalization group (RG). This allows us to obtain the decay rate for any renormalization scale and top-quark mass scheme and, in the postdiction of the last known coefficient, this strategy turns out to be more accurate than applying the approximants directly to $\Gamma(H\to gg)$. Our predictions for the QCD $\beta$ function also benefit from the knowledge of its large-$n_f$ coefficients~\cite{Gracey:1996he}.

The structure of this paper is as follows. In Section~\ref{sec:higgsdecaysqcd}, we review the perturbative description of $H\to gg$ in the heavy-top limit, of the gluonium correlator, of the decoupling relation, of the Wilson coefficient $C_1$, of the $\beta$ function, and of $H\to b\bar b$. Section~\ref{sec:borelandlargebeta0} discusses the Borel transform and the large-$\beta_0$ limit. In Section~\ref{sec:pas}, we introduce the Padé and D-log Padé approximants, while Section~\ref{sec:method} describes our strategy, including the criteria used to select the approximants, and validates it through postdictions of the last known coefficients. In Section~\ref{sec:results} we present our predictions for the unknown coefficients, the resulting $\Gamma(H\to gg)$ and its residual scale dependence, and the updated estimate for $H\to b\bar b$. Finally, Section~\ref{sec:conclusions} contains our conclusions. A few longer formulae are relegated to Appendices \ref{app:impig2beta0} and~\ref{app:RGeq} and detailed numerical results are given in Appendices \ref{app:results} and~\ref{app:fits}.

\section{Perturbative series}\label{sec:higgsdecaysqcd}

\subsection{Higgs boson decay to gluons}
Within the framework of the heavy-top limit, the effective Lagrangian that governs the Higgs decays to gluons through the effective $Hgg$ coupling can be written in terms of the renormalized gluonic fields and the QCD Lagrangian for $n_f$ light-quark flavors as
\begin{equation}
\mathscr{L}_{\text{eff}}^{(Hgg)} = \mathscr{L}_{\textrm{QCD}}(n_{f}) - \frac{H}{v} \left( C_{1} \,G_{\mu\nu}^{a}G^{\mu\nu}_{a}\right).
\label{eq:leffhgg}
\end{equation}
In this expression, $v$ is the vacuum expectation value of the Higgs field and $n_f$ the number of light quark flavors. The Wilson coefficient
$C_1\equiv C_1(\mu^2/m_t^2,\alpha_s(\mu^2))$ is responsible for the effective coupling between the Higgs boson and the gluons and governs the decay $H\to gg$ as well as the production of a Higgs boson via gluon fusion (the main production process at the LHC).
The higher-order QCD corrections for this process are calculated as an expansion in powers of $m_H^2/(4m_t^2)$, where the dominant term corresponds to the limit $m_t\to\infty$~\cite{Inami:1982xt,Chetyrkin:1997iv}.
In perturbative QCD, the decay width $\Gamma(H \rightarrow gg)$ begins at $\alpha_{s}^{2}$ and is currently known up to N4LO or $\mathcal{O}(\alpha_s^6)$ in the heavy-top limit~\cite{Herzog:2017dtz}.

The decay width for $H\to gg$ receives contributions from the Wilson coefficient $C_1$ associated with the effective $Hgg$ coupling of Eq.~(\ref{eq:leffhgg}), as well as the interactions between the gluons in the final state, governed by the so-called scalar gluonium correlator. Here we will study these quantities separately and start by discussing the gluonium correlator, which turns out to dominate the correction to $H\to gg$ at orders $\alpha_s^4$ and $\alpha_s^5$. The correlator can be defined as
\begin{equation}
\Pi_{G^{2}}(p^{2}) = i \int d^{4}x \,\, e^{ip \cdot x} \langle{\Omega| T\{j_G(x)j_G(0)\}|\Omega\rangle},
\label{eq:ggcorrelator}
\end{equation}
where $| \Omega \rangle$ is the non-perturbative QCD vacuum and $j_G(x)=G_{\mu\nu}^a(x)G^{\mu\nu \, a}(x)$ is the gluonium current. The gluonium correlator $\Pi_{G^2}$ is not directly related to any physical observable, but plays an important conceptual role in QCD --- due to its contribution to the anomalous trace of the
QCD energy-momentum tensor~\cite{Jamin:2012nn} --- and its imaginary part intervenes in the decay rate of $H \to gg$, computed at $p^2=s=m_H^2$, through~\cite{Herzog:2017dtz}
\begin{equation}
\Gamma(H \rightarrow gg) = \frac{\sqrt{2} \, G_{F}}{m_H}\, |C_{1}|^{2}\, \operatorname{Im} \Pi_{G^{2}}\left(s+ i 0\right),
\label{eq:hggd}
\end{equation}
where $G_F$ is the Fermi constant.

The imaginary part of $\Pi_{G^2}(s)$ has the following perturbative expansion~\cite{Jamin:2012nn}
\begin{equation}
\operatorname{Im} \Pi_{G^2} (s) = \dfrac{2}{\pi} \, s^2 \left[1 + \sum_{n=1}^\infty \, \alpha_s^n \sum_{k=1}^{n+1} g_{n,k} \, \ln^{k-1}\left(\frac{s}{\mu^2}\right) \right].
\label{eq:imgluoniumcorrelator}
\end{equation}
With the convenient choice $\mu^2=s=m_H^2$, which resums the logarithms associated with the running of $\alpha_s$, the independent coefficients $g_{n,1}$ can be expanded in powers of $n_f$ as
\begin{equation}
g_{n,1}= \sum_{k=0}^n g_{n,1}^{(k)} \,\, n_{f}^k,
\label{eq:gn}
\end{equation}
with the following numerical results (obtained from the analytic expressions of Ref.~\cite{Herzog:2017dtz})
\begin{align}
&g_{1,1} = 5.8092 - 0.37136 \, n_f, \nonumber \\
&g_{2,1} = 24.618 - 3.9894 \, n_f + 0.091368 \, n_f^2, \nonumber \\
&g_{3,1} = 82.339 - 24.899 \, n_f + 1.4989 \, n_f^2 - 0.017346 n_f^3, \nonumber \\
&g_{4,1} = 218.615 - 112.977 \, n_f + 12.7653 \, n_f^2 - 0.39786 \, n_f^3 + 0.0025809 \, n_f^4.
\label{eq:imgluoniumnfdependence}
\end{align}
Here, and throughout this paper, $n_f$ refers to the number of active light-quark flavors (which often appears as $n_\ell$ in the literature).
Note that here, since two gluon lines exist at leading order, in a term of $\mathcal{O}(\alpha_s^n)$, the highest power of $n_f$ is $n_f^n$. The coefficient of the highest power of $n_f$ for each $g_{n,1}$ coefficient is known to all orders from the large-$n_f$ results of Ref.~\cite{Jamin:2012nn}, with analytic expressions given in App.~\ref{app:impig2beta0}. The numerical values of $g_{n,1}^{(k)}$ can be read from Eq.~(\ref{eq:imgluoniumnfdependence}). Note that, for a given power of $\alpha_s$, these coefficients obey a very distinctive pattern: they display a sign alternation and a suppression in absolute value order by order in powers of $n_f$. This is quite general (although not universal) in QCD and can be qualitatively explained by the large-$\beta_0$ limit, as we discuss in the remainder of this paper. We will take advantage of this decomposition in powers of $n_f$ in order to better estimate the unknown coefficients of the series we are interested in.

We turn now to the Wilson coefficient $C_1$, which also intervenes in Higgs production through gluon-gluon fusion.
The standard method to determine $C_1$ is through a low-energy theorem~\cite{Chetyrkin:1997un,Baikov:2016tgj}. The coefficient can be derived from the decoupling relation, which connects the strong coupling value $\alpha_s^{(n_f)}$ in a theory with $n_f$ light flavors to the coupling with $n_f+1$ flavors, $\alpha_s^\prime\equiv \alpha _s^{(n_f+1)}$, and reads
\begin{equation}
\alpha_s^{(n_f)}(\mu) = \zeta_g^2(\mu, m_Q) \, \alpha_s^{(n_f +1)}(\mu),\label{eq:decouplingrelation}
\end{equation}
where the `decoupling' constant $\zeta_g^2$ is given in terms of an expansion in powers of $a_s^\prime\equiv \alpha_s^\prime/\pi$ as
\begin{equation}
\zeta_g^2 = 1 + \sum_{n=1}^{\infty} (a_s^{\prime})^n\sum_{k=0}^{n} d_{n,k} \, L_Q^k ,
\label{eq:decoupling}
\end{equation}
with $L_Q = \ln\left(\mu^2/m_Q^2(\mu^2) \right)$, where $m_Q(\mu^2)$ is the heavy-quark $\overline{\rm MS}$ mass at the scale $\mu^2$. Concretely, here, $m_Q$ is going to be the mass of the top quark, $m_t$.
The non-trivial independent coefficients are the $d_{n,0}$, while the $d_{n,k>0}$ can be written in terms of the lower-order $d_{n,0}$ as well as $\beta$- and $\gamma$-function coefficients, with the use of the RG. The relevant renormalization group equation (RGE) can be obtained taking $\mu \, d/d\mu$ on both sides of Eq.~(\ref{eq:decouplingrelation}), leading to
\begin{equation}
\left[ 2 + 2 \gamma^{\prime}_m(a_s^\prime) \right] \, \dfrac{\partial \zeta_g^2}{\partial L_Q} = - \dfrac{\beta(\zeta_g^2 \, a_s^\prime)}{a_s^\prime} + \zeta_g^2 \, \dfrac{\beta^\prime(a_s^\prime)}{a_s^\prime} + \beta^\prime(a_s^\prime) \, \dfrac{\partial \zeta_g^2}{\partial a_s^\prime},\label{eq:zetagRGE}
\end{equation}
where $\gamma^\prime_m$ and $\beta^\prime$ are the quark-mass anomalous dimension and the $\beta$-function, respectively, in the theory with $n_f$+1 quark flavors. From this relation, it is possible to determine the coefficients $d_{n,k>0}$ of $\zeta_g^2$ recursively in terms of $\beta$- and $\gamma_m$-function coefficients as well as lower-order $d_{n,0}$. One has, for example (using $d_{1,0}=0$)
\begin{align}
&d_{1,1} =-\frac{1}{2} \, \left( \beta_1 - \beta'_1\right) = -\frac{1}{6}, \nonumber\\
&d_{2,1} =-\frac{1}{2} \, (\beta_2 - \beta_2') - \gamma_{m,1} \, d_{1,1} =-\frac{11}{24},
\nonumber\\ &d_{2,2} = d_{1,1}^2 =\frac{1}{36},
\end{align}
in agreement with the results found in the literature~\cite{Spira:2016zna,Chetyrkin:2016uhw}. The first four independent $d_{n,0}$ coefficients are known analytically, with the following numerical results~\cite{Baikov:2016tgj}
\begin{align}
d_{1,0} &= 0 , \nonumber \\
d_{2,0} &= 0.152778, \nonumber \\
d_{3,0} &= 0.972057 - 0.0846515 \, n_f, \nonumber \\
d_{4,0} &= 5.17035 - 1.00993 \, n_f - 0.0219784 \, n_f^2 .
\label{eq:nfdependencedecoupling}
\end{align}

The low-energy theorem then gives the Wilson coefficient $C_1$ as
\begin{equation}
C_1 = \frac{1}{2}\frac{\partial }{\partial L_Q} \, \ln\zeta_g^2.
\label{eq:LET}
\end{equation}
From the expansion of $\zeta_g^2$ one then gets $C_1$, including the powers of $L_Q$. In this case, because of the derivative, the coefficients of $C_1$ will contain $d_{n,k>0}$ coefficients even when the logarithms are resummed.

The Wilson coefficient can be expanded perturbatively in powers of $a_s\equiv a_s^{(n_f)}$ as
\begin{equation}
C_{1} = -\frac{1}{12} \,\, a_s \left( 1 + \sum_{n=1} c_{n} \, a_s^n(\mu) \right).
\label{eq:C1def}
\end{equation}
In the concrete case of $H\to gg$, $n_f=5$ and $n_f+1=6$.
This series is currently known up to the fourth order~\cite{Chetyrkin:2016uhw,Herzog:2017dtz}.\footnote{In Ref.~\cite{Chetyrkin:2016uhw} the perturbative expansion is done in terms of $\alpha_s^{(n_f+1)}$ instead of $\alpha_s^{(n_f)}$ as we do here.} The coefficients $c_n$ can also be expanded in powers of $n_f$ and we write
\beq
c_n = \sum_{m=0}^{n-1} c_n^{(m)} \, n_f^m , \label{eq:c1-coeff}
\eeq
where the numerical values for the coefficients $c_n^{(m)}$ in the $\overline{\mathrm{MS}}$ scheme are given by~\cite{Spira:2016zna}
\begin{align}
c_1 &= 2.75 , \nonumber\\
c_2 &= 9.6424 + 1.1875 \, L_t - \left( 0.69792 - 0.33333 \, L_t \right) \, n_f , \nonumber \\
c_3 &= 47.370 + 6.0174 \, L_t + 3.2656 \, L_t^2 - \left( 7.6897 - 1.0185 \, L_t - 0.71875 \, L_t^2 \right) \, n_f \, \nonumber\\
& - \left( 0.22071 - 0.044560 \, L_t + 0.055556 \, L_t^2 \right) \, n_f^2 , \nonumber \\
c_4 &= 311.78 + 26.504 \, L_t + 21.878 \, L_t^2 + 8.9805 \, L_t^3 \, \nonumber \\
&- \left( 62.368 + 10.212 \, L_t - 1.7964 \, L_t^2 - 1.4323 \, L_t^3 \right) \, n_f \, \nonumber \\
&+ \left( 1.6155 - 2.4258 \, L_t - 0.091652 \, L_t^2 - 0.27257 \, L_t^3 \right) \, n_f^2 \, \nonumber \\
&- \left( 0.034339 - 0.092786 \, L_t + 0.011140 \, L_t^2 - 0.0092593 \, L_t^3 \right) \, n_f^3.
\end{align}
In these expressions, the logarithms are $L_t = \ln{(\mu^2/m_t^2(\mu^2))}$, with the top-quark mass evaluated in the $\overline{\mathrm{MS}}$ scheme at the renormalization scale $\mu^2$, and with the running performed in the theory with $n_f+1$ flavors. Analytic expressions for $c_4$ and $c_5$ in terms of $d_{n,k}$ and $\beta$-function coefficients are given in App.~\ref{app:RGeq}. For the $c_i$ coefficients, the highest power of $n_f$ for a coefficient of order $n$ is $n_f^{n-1}$. We are particularly interested in the first unknown coefficient $c_5$, which we will predict here, as a function of $d_{5,0}$. As shown in Eq.~(\ref{eq:c5}) in App.~\ref{app:RGeq}, this coefficient also requires the unknown six-loop coefficient of the $\beta$ function, which we denote $\beta_6$ [see Eq.~(\ref{eq:betafunc-def})]. We will, therefore, study $\beta_6$ as well, as a function of the powers of $n_f$, in order to be able to obtain the coefficient~$c_5$.

The decay width for Higgs into gluons can then be expanded perturbatively, using the $g_{n,1}$ and $c_n$ coefficients in Eq.~(\ref{eq:hggd}) as
\begin{equation}
\Gamma(H \rightarrow gg) = \dfrac{m_H^3 \, \alpha_s^2}{72 \pi^3 v^2} \left[ 1 + \sum_{n =1}^{\infty} \gamma_{n} \,\alpha_{s}^n \right].
\label{eq:decayrateHgg}
\end{equation}
The series coefficients, $\gamma_n$, of $\Gamma(H \rightarrow gg)$ depend on the scheme chosen for the top-quark mass when dealing with the logs of $C_1$. In the $\overline{\mathrm{MS}}$ scheme with the choice $\mu^2 = m_H^2$, which resums the logarithms of
$\operatorname{Im} \Pi_{G^2}$ in Eq.~(\ref{eq:imgluoniumcorrelator}),
the dependence on $n_f$ of the known coefficients $\gamma_n$, up to order $\alpha_s^6$, is given by~\cite{Spira:2016zna,Herzog:2017dtz}
\begin{align}
&\gamma_1 = 7.5599 - 0.37136 \, n_{f}, \nonumber \\
&\gamma_2 = 37.373 - 4.8191 \, n_f + 0.091368 \, n_f^2, \nonumber \\
&\gamma_3 = 144.95 - 34.561 \, n_f + 1.7085 \, n_f^2 - 0.017346 \, n_f^3, \nonumber \\
& \gamma_4 = 462.08 - 178.66 \, n_f + 16.549 \, n_f^2 - 0.44021 \, n_f^3 + 0.0025809 \, n_f^4,
\label{eq:hggnfdependence}
\end{align}
with the following numerical values for $n_f = 5$
\begin{equation}
\gamma_1 = 5.7031, \qquad \gamma_2 = 15.562, \qquad \gamma_3 = 12.686, \qquad \gamma_4 = -70.938,\label{eq:g4-exact}
\end{equation}
where, when evaluating $L_t$ at $\mu^2=m_H^2$, we used $m_H = 125.13$~GeV and $m_t(m_t^2) = 162.5$~GeV.
The coefficient of the highest power of $n_f$ for each $\gamma_n$ coincides with that of the respective $g_{n,1}$ coefficient, as can be seen comparing with Eq.~(\ref{eq:imgluoniumnfdependence}) and is therefore known to all orders in $\alpha_s$ from the results of Ref.~\cite{Jamin:2012nn}. We will take advantage of this knowledge in the remainder of the paper.
Furthermore, the value of $\gamma_4$ turns out to be dominated by the contribution from $g_{4,1}$.
For $\mu^2=m_H^2$ one finds, approximately, the following contributions to $\gamma_4$ with $n_f=5$
\beq
\gamma_4= 2.215-(75.257)_{g_{4,1}}+(0.278)_{d_{4,0}}+(1.829)_{\beta_5}, \label{eq:gamma4-components}
\eeq
where the contributions from $g_{4,1}$, $d_{4,0}$, and $\beta_5$ are identified explicitly.
This shows that, for a good determination of $\gamma_4$, the most important ingredient is a good determination of $g_{4,1}$.

As mentioned, in order to obtain the coefficient $c_5$, which is needed for an estimate of $\gamma_5$, we need an estimate for the six-loop coefficient of the $\beta$ function. Here, we define the $\beta$ function as~\cite{Beneke:2008ad}
\beq
-\mu \, \frac{d a_s}{d \mu} =\sum_{i=1} \beta_i \, a_s^{i+1} = \beta_1 \, a_s^2 + \beta_2 \, a_s^3 + \cdots . \label{eq:betafunc-def}
\eeq
The coefficients $\beta_i$ that have been computed exactly, for $N_c=3$, are
\begin{align}
&\beta_1 = \dfrac{11}{2} - \dfrac{1}{3} \, n_f, \nonumber\\[3pt]
&\beta_2 = \dfrac{51}{4} - \dfrac{19}{12} \, n_f, \nonumber\\
&\beta_3 = 44.641 - 8.7379 \, n_f + 0.18808 \, n_f^2, \nonumber \\
&\beta_4 = 228.46 - 54.268 \, n_f + 3.1648 \, n_f^2 + 0.011713 \, n_f^3, \nonumber \\
&\beta_5 = 1049.1 - 363.60 \, n_f + 34.312 \, n_f^2 - 0.45171 \, n_f^3 - 0.0035986 \, n_f^4.
\label{eq:betafunction}
\end{align}

\subsection{\boldmath \texorpdfstring{$H\to b\bar b$}{H to b bar b}}

In the context of the Standard Model, the Higgs boson predominantly decays into bottom quarks~\cite{Djouadi:2005gi}. Here we consider these decays as described by the following effective Lagrangian
\begin{equation}
\mathscr{L}_{\text{eff}}^{(Hb\bar b)} = \mathscr{L}_{\textrm{QCD}}(n_{f}) - \frac{H}{v} \left( C_{2}\, m_b \, \bar b b \right),
\label{eq:leffhbb}
\end{equation}
but with $C_2=1$; we take into account, therefore, the contributions often referred to as the `massless contributions'~\cite{Davies:2017xsp} even though they are proportional to an overall $m_b$ factor.
In this set-up,
the decay rate $\Gamma(H \to b \bar{b})$ is described in terms of the correlator of two scalar currents, which is defined as
\begin{equation}
\Pi (p^2) \equiv i \int d^4x \, \mathrm{e}^{ip\cdot x} \, \langle \Omega | T\{j(x)j^\dag(0)\} | \Omega \rangle ,
\end{equation}
with $j(x) = m_q \, \bar{q}(x) \, q(x)$. In the limit where the bottom quark is massless, which is the leading result in an expansion in powers of $m_b^2/m_H^2$, the quark fields are taken to be massless in the quark loops, which reduces the number of mass scales and simplifies the loop integrations. In this limit, the QCD corrections to the scalar correlator are known up to $\mathcal{O}(\alpha_s^4)$~\cite{Baikov:2005rw,Herzog:2017dtz}.

The general form for the purely perturbative expansion of the massless scalar correlator $\Pi(s)$ in powers of the strong coupling is given by
\begin{equation}
\Pi(s) = - \dfrac{N_c}{8\pi^2} \, m_q^2(\mu^2) \, s \sum_{n=0}^\infty \alpha_s^n(\mu) \sum_{k=0}^{n+1} h_{n,k} \, L^k ,
\label{eq:scalarcorrelator}
\end{equation}
with $\mu$ being the renormalization scale and $L \equiv \ln{(-s/\mu^2)}$. The coefficients $h_{n,0}$ do not contribute to any physical quantity and are related to the renormalization scheme prescription. The presence of $h_{n,0}$ indicates that the correlator itself is not a physical quantity in the RG sense. The independent coefficients are the $h_{n,1}$, since the other $h_{n,k>1}$ can all be obtained through RGEs~\cite{Chetyrkin:1996sr,Jamin:2021qxb,Jamin:2016ihy}.

In our case of interest, i.e. $H \to b \bar{b}$, the decay rate is proportional to the imaginary part of the scalar correlator as
\begin{equation}
\Gamma(H \rightarrow b\bar{b}) = \dfrac{1}{v^2 \, m_H} \, \operatorname{Im} \Pi(s+i0).
\end{equation}
The imaginary part of $\Pi(s)$ is a physical observable, thus it satisfies a homogeneous RGE and the logarithms can be resummed with $\mu^2 = s = m_H^2$. In this case, we have that the perturbative expansion of $\operatorname{Im} \Pi(s)$ is~\cite{Baikov:2005rw,Herzog:2017dtz}
\begin{equation}
\operatorname{Im} \Pi (s) = \dfrac{N_c}{8\pi} \, m_q^2(m_H^2) \, s \sum_{n=0}^\infty \alpha_s^n(m_H) \sum_{k=0}^{[n/2]} h_{n,2k+1} \, (i\pi)^{2k} = \dfrac{N_c}{8\pi} \, m_q^2 \, s \left[1 + \sum_{n=1}^\infty p_n \, \alpha_s^n \right],
\label{eq:imscalarcorrelator}
\end{equation}
where $[x]$ represents the greatest integer not greater than $x$. The coefficients $p_n$ up to fourth order in terms of $n_f$ are given, numerically, by
\begin{align}
&p_1 = 1.8038, \nonumber \\
&p_2 = 3.6415 - 0.13766 \, n_f, \nonumber \\
&p_3 = 5.2937 - 0.83116 \, n_f + 0.008352 \, n_f^2, \nonumber \\
&p_4 = 0.40383 - 2.2681 \, n_f + 0.099424 \, n_f^2 - 0.000210041 \, n_f^3,
\label{eq:hbbnfdependence}
\end{align}
which, for $n_f = 5$, give
\begin{equation}
p_1 = 1.8038, \qquad p_2 = 2.9532, \qquad p_3 = 1.3468, \qquad p_4 = - 8.4771.
\end{equation}

The second derivative of the scalar correlator with respect to $s$ is also associated with an observable in the RG sense~\cite{Jamin:2021qxb,Jamin:2016ihy}, because it removes the two scheme-dependent constants of $\Pi(s)$ and will be very important for our analysis, as we will see. Its perturbative expansion in the Euclidean region, $s=-Q^2<0$, with $\mu^2=Q^2=m_H^2$ is
\begin{align}
&\Pi^{''}(s) = - \dfrac{N_c}{8\pi^2} \, \dfrac{m^2_q(m_H^2)}{s} \left[ 1 + \sum_{n=1}^\infty k_n \, \alpha_s^n(m_H) \right] \nonumber \\
&= - \dfrac{N_c}{8\pi^2} \, \dfrac{m^2_q(m_H^2)}{s} \left[ 1 + 1.1671 \, \alpha_s + 1.2979 \, \alpha_s^2 + 1.2799 \, \alpha_s^3 + 1.5805 \, \alpha_s^4 + \dots \right] ,
\label{eq:secderscalarcorrelator}
\end{align}
where the numerical values are computed considering five active quark flavors. The coefficients $k_n$ can be written as a function of the coefficients $h_{n,k}$ of the scalar correlator as $k_n = h_{n,1} + 2 h_{n,2}$. Since the second derivative is also a physical quantity, it also satisfies a homogeneous RGE, which one can use to reinstate the logarithmic terms.

\section{Borel transform and large-\texorpdfstring{$\boldsymbol{\beta_0}$}{beta0} limit}
\label{sec:borelandlargebeta0}

The perturbative series described in Sec.~\ref{sec:higgsdecaysqcd} are expected to be factorially divergent, but asymptotic. A useful approach to deal with them is the application of the Borel transform. Consider a function $f$ with a factorially divergent power series of the form
\begin{equation}
f(\alpha_{s}) = f(0) + \sum_{n=1}^{\infty}f_{n}\,\alpha_{s}^{n},
\label{eq:aforem}
\end{equation}
which has a zero convergence radius. We assume that the $f_n$ display the decomposition in powers of $n_f$
\beq
f_n(n_f) = \sum_{k=0}^{m} f_n^{(k)} \, n_f^k.
\label{eq:fnm-coeff}
\eeq
Here $m$ is the highest power of $n_f$ in the coefficient $f_n$ (which is often $m=n-1$ or $m=n$) and we made explicit the $n_f$ dependence in $f_n(n_f)$, as we will often do in the remainder.
The Borel transform for the $\alpha_s$-dependent part of this series, $B[f]$, can be defined as follows
\begin{equation}
B[f](u) = \sum_{n=0}^{\infty} \frac{f_{n+1}}{n!} \left( \frac{u}{\beta_{0}} \right)^{n},
\label{eq:boreltransform}
\end{equation}
where $\beta_0=\beta_1/(2\pi)$, and $\beta_1$ is the first coefficient of the QCD $\beta$ function. In the $u$ plane, the factorial growth of coefficients is mitigated by the presence of $n!$ in the denominator. A representation for the original function can be obtained from the Borel transform through the inverse transformation, which is essentially a Laplace transform of $B[f]$, written as
\begin{equation}
\tilde{f}(\alpha_{s}) = \frac{1}{\beta_{0}} \int_{0}^{\infty}du \,\,\, e^{-u/(\beta_{0}\alpha_{s})} \, B[f](u),
\label{eq:laplace}
\end{equation}
with $\tilde f(\alpha_s)=f(\alpha_s)-f(0)$.
The series of Eq.~(\ref{eq:aforem}) can belong to two categories. Borel summable series have a well-defined integral without singularities on the positive real axis, yielding a unique sum. In contrast, series with singularities along the integration path require a prescription to circumvent them, which introduces ambiguities. However, power corrections in the Operator Product Expansion (OPE) are, on general grounds, expected to cancel these ambiguities~\cite{Beneke:1998ui}.
In the context of the present work, the Borel transform is the basis of the Pad\'e-Borel method, i.e. the construction of Pad\'e approximants to the Borel transformed series as a means of accelerating the convergence of the procedure~\cite{Baker:1976ff,Ellis:1995jv,Boito:2018rwt,Boito:2021scm}.

In our context, singularities in the Borel plane are the renormalons of perturbation theory. They are responsible for determining the asymptotic behavior of the series. In QCD, singularities on the negative real axis of $u$ are of UV origin, and they generate sign alternating series coefficients at high orders. Renormalons located on the positive real axis of $u$ are of IR origin, and lead to fixed-sign coefficients. In general, an infinite number of renormalons coexist, and the closest to the origin dominates the series behavior at intermediate and large orders.

Since renormalons are the result of the summation of infinitely many terms in the perturbative series, a lot of our knowledge about them is inferred from the large-$\beta_0$ limit,\footnote{In our notation, this limit should be referred to as the `large-$\beta_1$ limit.' However, since the term large-$\beta_0$ limit is conventionally used in the literature, we will retain this terminology throughout.} in which the $\alpha_s$ expansion is known to all orders. Large-$\beta_0$ results are obtained from the large-$n_f$ limit, where quark-bubble loop corrections to the gluon propagator are resummed to all orders.
These results exactly reproduce, therefore, the largest power of $n_f$ in the perturbative coefficients of the QCD series. In this work, we will use the knowledge about the highest power of $n_f$, obtained in the context of the large-$\beta_0$ expansion, as input in our analysis to estimate the higher-order coefficients in full QCD for $\Gamma(H \rightarrow gg)$, as well as when updating results for $\Gamma(H \rightarrow b \bar{b})$.

The perturbative coefficients in the large-$\beta_0$ limit are obtained from the large-$n_f$ results with the procedure known as naive non-Abelianization~\cite{Beneke:1994qe,Broadhurst:1994se}. It consists in replacing the fermionic contribution to the leading $\beta$-function coefficient, $\beta_{1f}=-2T_F \, n_f/3=-n_f/3$, with the full coefficient $\beta_1=11C_A/6-2 T_F \, n_f/3$, thereby including non-Abelian contributions.\footnote{For $SU(3)$, $C_A=3$ and $T_F=1/2$.} In terms of $n_f$, this amounts to the replacement $n_f\to-3\beta_1 = n_f-33/2$. This means that the perturbative coefficients in the large-$\beta_0$ limit are generated from the largest power of $n_f$ in Eq.~(\ref{eq:fnm-coeff}). If the largest power of $n_f$ is given by $m$ then
\beq
f_n^{[\mathrm{L}\beta_0]}= f_n^{(m)} \left(-3\beta_1\right)^{m} = f_n^{(m)}\sum_{k=0}^{m}\binom{m}{k} (-1)^{m-k}\left(\frac{33}{2}\right)^{m-k} n_f^k.
\eeq
This predicts that the $n_f$ coefficients will show a strict sign alternation with higher powers of $n_f$ being suppressed by powers of $\sim 2/33$ (modulated by binomial coefficients) --- which reproduces, in the majority of cases, the behavior of the exact pQCD series we have shown in the previous section.

Although the result for $\Gamma(H \rightarrow gg)$ in the large-$\beta_0$ limit is not found in the literature, we point out here that sufficient information exists to easily obtain it. From Eqs.~(\ref{eq:gn}) and~(\ref{eq:c1-coeff}) it is clear that the leading $n_f$ power in the pQCD expansion of $\Gamma(H \rightarrow gg)$ is entirely driven by the expansion of the gluonium correlator, through the $g_{n,1}$ coefficients, which have been known in the large-$\beta_0$ limit since 2012~\cite{Jamin:2012nn}. This happens because for the $g_{n,1}$ coefficients the highest power of $n_f$ is $n_f^n$ (and not $n_f^{n-1}$) which is a consequence of the fact that the leading order already contains gluon propagators that can be dressed with quark-bubble loops, each of which counts as $\alpha_s n_f$. A clear shortcoming in this specific case is that, in the large-$\beta_0$ limit, the coefficients $\gamma_n^{[{\rm L}\beta_0]}$ and $g_{n,1}^{[{\rm L}\beta_0]}$ are the same, without reference to $C_1$. This is particularly important for lower order $\gamma_n$, while the result for $\gamma_4$ shows a much smaller contribution from $C_1$. The large-$\beta_0$ result for $g_{5,1}$ can be extracted from the analytic expressions of Ref.~\cite{Jamin:2012nn} which gives, for the highest power of $n_f$, $g_{5,1}^{(5)}=-0.0002821$ and, for $g_{5,1}^{[{\rm L}\beta_0]}$, in powers of $n_f$,
\begin{equation}
g_{5,1}^{[{\rm L}\beta_0]} = 344.977 - 104.539 \, n_{f} + 12.6713 \, n_{f}^2 - 0.76796 \, n_f^3 + 0.023272 \, n_f^4 - 0.0002821 \, n_f^5.
\label{eq:estimateB0}
\end{equation}
Again, the coefficient of the highest power of $n_f$ is the same as in full QCD and will be used in order to constrain the prediction of the first unknown QCD coefficient. Analytic results for the coefficient of the highest power of $n_f$ in the $g_{n,1}$ coefficients can be found in App.~\ref{app:impig2beta0}.

\section{\texorpdfstring{Pad\'e}{Padé} and D-log \texorpdfstring{Pad\'e}{Padé} approximants}\label{sec:pas}
To make predictions for higher-order coefficients we will use Pad\'e approximants (or simply PAs) and some of their variants. The approximant $P^{M}_{N}(z)$ is obtained through the ratio of two polynomials of order $M$ and $N$ as follows
\begin{equation}
P_{N}^{M}(z) = \frac{Q_{M}(z)}{R_{N}(z)} = \frac{a_{0} + a_{1}z + \cdots + a_{M}z^{M} }{1 + b_{1}z + \cdots + b_{N}z^{N}},
\label{eq:PA}
\end{equation}
assuming, without loss of generality, that $R_N(0)=1.$
This rational approximant is said to `make contact' of order $M+N$ with the Taylor series expansion of a function in the complex plane around $z= 0$. This means that the coefficients of the PA to a given function are obtained by expanding Eq.~(\ref{eq:PA}) in powers of $z$ and matching to the first $N+M+1$ terms of the exact Taylor series of the original function. Thus, a PA of order $N+M$, $P_N^M(z)$, has $(N+M+1)$ parameters and replicates the first $(N+M+1)$ coefficients of the Taylor series expansion of the original function. The re-expansion of Eq.~(\ref{eq:PA}) provides then estimates for the coefficients of order $N+M+1$ and higher.

Besides the Taylor series coefficients, the PAs also reproduce, approximately, the analytic structure of the function of interest by reproducing the poles and their residues, which converge to their exact values as the order of the PA grows. However, this is done in a hierarchical way: the poles closer to the origin are already reproduced by lower order PAs, while poles that lie further away require higher-order PAs before they appear in the result.

In Pad\'e theory, there are convergence theorems for PAs when applied to meromorphic functions (whose singularities are isolated poles). Pommerenke's theorem, for instance, ensures that PAs of the sequence $P_N^{N+k}$, with $k \geq -1$, will converge to the original function~\cite{Baker1996pade,POMMERENKE1973775, baker2000defects} on compact sets minus an exceptional set of arbitrarily small measure. Furthermore, this theorem guarantees that extraneous poles, i.e. the ones that cannot be related to any singularity of the original function, will move far from a compact region in the complex plane with the increase of the order of the PA or they will appear in pairs, with a related zero located very closely to the pole. The latter type of poles is known as \textit{defects} or \textit{Froissart doublets} and their appearance has an important consequence for the convergence of PA sequences: since the pole is almost exactly canceled by a close-by zero, the order of the PA is effectively reduced by two, disrupting the expected convergence pattern of a PA sequence. Since those poles are transient and do not accumulate anywhere in particular, higher order PAs resume the expected convergence (as shown in practice, e.g., in Ref.~\cite{Boito:2018rwt}). It is important to mention that close to these extraneous poles the PAs fail, but one can still use these approximants to study the desired function away from these singularities.

Here, we will build PAs not only to the series in $\alpha_s$, but also to the Borel transformed series, using the Pad\'e-Borel method~\cite{Baker:1976ff}.
The Borel transforms of the perturbative series we are working with have superimposed cuts in full QCD, not isolated poles. In the particular case of Stieltjes functions, specific PA sequences converge to the function even in the presence of a cut in the complex plane. The cut is then mimicked by the accumulation of poles along the cut (interleaved with zeros). Experience shows that PAs can also reproduce cuts of functions that are not Stieltjes in the same way~\cite{Baker1996pade,MasjuanQueralt:2010hav,Costin:2021bay}.

An additional tool to deal with functions that have cuts is to employ D-log Pad\'e approximants~\cite{Baker1996pade}, which we will denote as D-log PAs. For instance, consider a function of the type
\begin{equation}
f(z) = \frac{A(z)}{(p - z)^{\rho}} + B(z),
\end{equation}
where $A(z)$ and $B(z)$ are analytic functions at $z = p$. Instead of applying Padé approximants to the function $f(z)$, it is convenient to work with the logarithmic derivative of $f(z)$ which, for $z$ close to $p$, displays a simplified singularity structure (up to corrections that are regular at $z=p$)
\begin{equation}
F(z) = \frac{\mathrm{d}}{\mathrm{d}z} \, \ln [f(z)] \sim \frac{\rho}{p - z},
\label{eq:FDlog}
\end{equation}
and is more amenable to the treatment with usual PAs.
The PA applied to the function $F(z)$, which we denote $\bar{P}^{M}_{N}(z)$, can then be used to provide an approximation for the function $f(z)$; this approximation is not necessarily rational. Here, we denote these approximants as $\text{Dlog}^{M}_{N} (z)$, having the following structure
\begin{equation}
\text{Dlog}^{M}_{N}(z) = f(0) \, \, \text{exp}\left[\int_0^z \mathrm{d}z^\prime \, \bar{P}^{M}_{N}(z^\prime)\right],
\end{equation}
where the normalization constant $f(0)$ must be reintroduced due to the loss of a constant in the differentiation process. The D-log PA requires the knowledge of the first $(N+M+2)$ Taylor series coefficients of $f(z)$, and the higher-order coefficients are predicted by the approximant. The ${\rm Dlog}_N^M$ requires, thus, one more input coefficient than $P_N^M$.

\section{Strategy and validation}\label{sec:method}

The strategy that we employ here for the reconstruction of higher orders of the perturbative series from rational approximants is based on previous works co-authored by some of us~\cite{Boito:2018rwt,Boito:2021scm}, but here we exploit the $n_f$ dependence of the coefficients in a way similar to what was done long ago~\cite{Ellis:1995jv,Ellis:1996zn,Ellis:1997sb}, and use systematically the exact large-$n_f$ results to impose constraints on the PAs. Previous experience, in part based on the systematic application of PAs to results in the large-$\beta_0$ limit, where the series is known to all orders, shows that the method is reliable~\cite{Boito:2018rwt,Boito:2021scm} but, given the scarce information we have in realistic cases, some of the PAs can be outliers and here we use the exact large-$n_f$ results, when available, to guide us in identifying and discarding some of these problematic approximants. Our main guiding principle is to be as model independent as possible --- without imposing constraints from assumptions about renormalons, for example, and imposing only exact large-$n_f$ results as constraints.

The method we employ is as follows. We build all possible PAs and D-log PAs to the series in $\alpha_s$ and to its Borel transform, with $n_f=1,...,9$, given the amount of information we have available, i.e. the number of coefficients that we know exactly (three or four, typically), with the exception of PAs of the type $P^0_N$.\footnote{The reason for that is that $P^0_N$ do not have any structure in the numerator and are forced to adjust residues of poles by placing transient poles on the real axis, often very far from the origin, effectively reducing the order of the PA.} These results are obviously not independent, but the series have different singularity structures and, therefore, the ability of different approximants to reproduce the coefficients can vary substantially. We then filter the approximants to exclude potentially problematic results from our final analysis discarding, for example, those that display a Froissart doublet, which effectively reduces the order of the PA by two. The
criteria for the filtering of the approximants are the following.
{\it(i)} We exclude the approximants that have a Froissart doublet, characterized by a pole close to a zero. This is quantified by the normalized residue of such a pole, $R_{\mathrm{norm}}$, defined as
\begin{equation}
R_{\mathrm{norm}} = \frac{|R|}{\mathcal{F}_0 \, |p|},
\end{equation}
where $R$ is the residue of the pole, $p$ is the pole position in the PA variable, with $\mathcal{F}_0=f(0)$ for PAs built to the series in $\alpha_s$, and $\mathcal{F}_0=B[f](0)$ for PAs built to the Borel transform of the series. This
produces a result that is comparable across different series and different approximants. If $R_{\rm norm}<10^{-2}$, the approximant is excluded from the final analysis. In the cases where $\mathcal{F}_0=0$, such as PAs built to the $\beta$ function in $\alpha_s$, $R_{\mathrm{norm}}$ is normalized only by the pole position.
{\it (ii)} For similar reasons, we exclude approximants that have a real pole far from the origin whose relative contribution $w_k$ to the predicted coefficient does not exceed $10^{-5}$. The contribution of the $k$-th pole of the approximant, at position $p_k$, to the predicted coefficient of order $n$ is proportional to $R_k$/$p_k^{n+1}$. We therefore define
\begin{equation}
w_k = \frac{|R_k/p_k^{n+1}|}{\sum_{j=1}^N |R_j/p_j^{n+1}|},
\end{equation}
where $N$ is the order of the approximant's denominator, $n$ is the order of the predicted coefficient, $R_k$ is the residue of the $k$-th pole, and $p_k$ denotes its position. For the D-log approximants, this criterion is applied to the Pad\'e approximant $\bar{P}_N^M$ used to construct the D-log. In this case, however, $w_k$ quantifies the contribution to the coefficient of the series $F(z)$ defined in Eq.~(\ref{eq:FDlog}). It is important to mention that this criterion cannot be applied to approximants with only one pole.\footnote{In the specific case of $\mathrm{Dlog}^1_1$, poles far from the origin in the PA $\bar P_1^1$ used to build it typically have large residues, $R$, which then turn into a large exponent for the D-log singularity after deconvolution, since $\mathrm{Dlog}_1^1(z) \propto 1/(p-z)^R$. Therefore, we discard D-log Pad\'es obtained from $\bar P_1^1$ with a single pole that has a very large residue $R$.}
{\it (iii)} PAs and D-log PAs that have complex poles with a distance from the origin smaller than 0.2 are excluded. Such nearby complex singularities are not expected and are spurious, potentially leading to unreliable estimates for the higher-order coefficients.
These selection criteria are applied systematically to all PAs in all cases. We stress, however, that tests performed at higher orders using large-$\beta_0$ results support these cuts {\it on average}, i.e. outliers can exist and it may happen that a reasonable PA is discarded by the criteria that we described above.

A final criterion is applied to the PAs that pass the previous three cuts in order to exploit the exact knowledge from large-$n_f$ computations (when available). We perform a least squares fit to the predictions of each PA for the coefficient of interest for $n_f=1,...,9$, in order to determine the coefficients of its polynomial dependence on $n_f$, $f_n^{(k)}$, see Eq.~(\ref{eq:fnm-coeff}). The result for the coefficient of the highest power of $n_f$, $f_n^{(m)}$, obtained from this fit is then compared with exact large-$n_f$ results as
\beq
\Delta_{n_f} = \left| \frac{f_n^{(m)}\big|_{\rm PA}-f_n^{(m)}\big|_{{\rm large-}n_f}}{f_n^{(m)}\big|_{{\rm large-}n_f}} \right|. \label{eq:Deltanf}
\eeq
PAs that have $\Delta_{n_f}$ significantly larger than other PAs built to the same quantity are discarded. Discarding these PAs imposing the exact knowledge from the large-$n_f$ result often leads to significant improvement in the results, both in terms of central value as well as for the final uncertainties, which can be tested for the postdiction of the last known coefficient of the series of interest, as discussed below.

After filtering, we produce an average result for each value of $n_f$ with an uncertainty given by the standard deviation of the predictions for each $n_f$. These values are then fitted with a polynomial in $n_f$, imposing, whenever available, the large-$n_f$ result. Once the polynomial is fixed, we obtain the final values of the coefficients for different $n_f$ values taking into account the strong correlations existing in intermediate results.

We turn now to the postdiction of the last known coefficient for some of the series of interest in order to describe and validate the strategy in concrete cases. We start with the discussion of the postdiction of the coefficient $g_{4,1}$, the last known coefficient of ${\rm Im}\Pi_{G^2}(s)$, shown in Eq.~(\ref{eq:imgluoniumnfdependence}), which can be written as a polynomial in $n_f$ as
\begin{align}
g_{4,1}=\sum_{k=0}^4 g_{4,1}^{(k)} \, n_f^k = 218.62 - 112.98 \, n_f + 12.765 \, n_f^2 - 0.3979 \, n_f^3 + 0.0025809 \, n_f^4.
\label{eq:nfdependenceex}
\end{align}
We only build PAs for which $g_{4,1}$ is the first forecast coefficient, using as input the $g_{i<4,1}$. One can, for example, build the $P_2^1(\alpha_s)$ to the series in $\alpha_s$, which, for $n_f=5$, gives
\begin{equation}
P_2^1(\alpha_s) = \dfrac{1 - 0.0107485 \, \alpha_s}{1 - 3.96310 \, \alpha_s + 8.70802 \, \alpha_s^2} \, .
\end{equation}
Expanding it, we obtain a prediction for $g_{4,1}^{\mathrm{PA}}(5) = -87.72$. This result is off by less than $17\%$ when comparing it to the true value, $g_{4,1}(5) = -75.26$. We constructed this PA for $n_f=1,...,9$ and obtained estimated values $g_{4,1}(n_f)$, a sample of which can be seen in Tab.~\ref{tab:postdictiongluonium} in App.~\ref{app:results}. Results from other PAs and D-log PAs that pass our consistency criteria are also shown in Tab.~\ref{tab:postdictiongluonium}. PAs or D-log PAs denoted $P^M_N(\alpha_s)$ or $\mathrm{Dlog}^M_N(\alpha_s)$ are built to the series in powers of $\alpha_s$ while those denoted $P^M_N(u)$ are built to its Borel transform. Finally, the results from $P^1_1(u)$ and ${\rm Dlog}^0_1(u)$ are discarded since they lead to a large-$n_f$ term that is highly incompatible with the exact value, as per the test of Eq.~(\ref{eq:Deltanf}), shown in the last column of Tab.~\ref{tab:postdictiongluonium}. Discarding these PAs through the test of Eq.~(\ref{eq:Deltanf}) turns out to be important to achieve a good reproduction of the exact values.

\begin{figure}[t]
\centering
\includegraphics[width=0.4\textwidth]{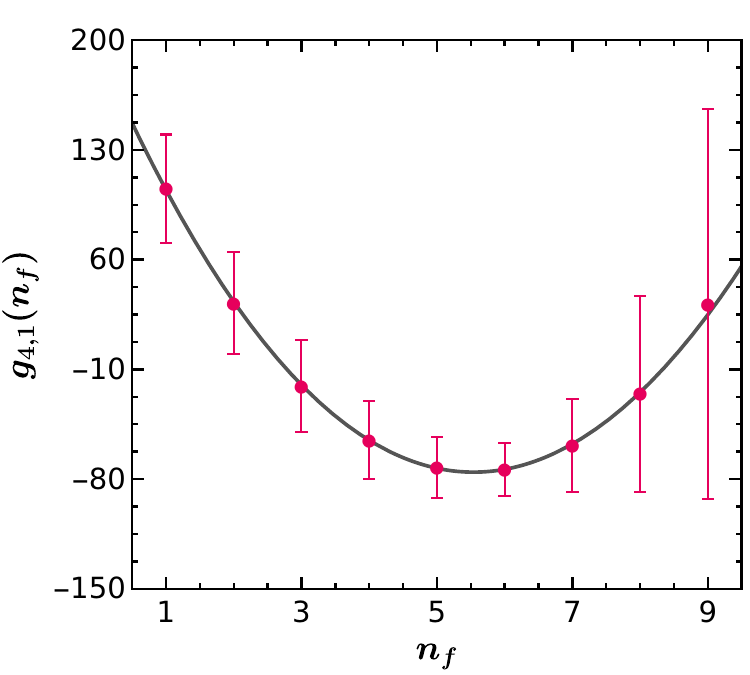}
\caption{Estimate of the coefficient $g_{4,1}(n_f)$ obtained through the average of the predictions from different approximants for various values of $n_f$. The polynomial fit of Eq.~(\ref{eq:g4pred}) is indicated by the gray line.}
\label{fig:g4}
\end{figure}

Results from three remaining PAs, $P^1_2(\alpha_s)$, ${\rm Dlog}^2_0(\alpha_s)$ and ${\rm Dlog}^1_0(u)$ are then averaged for each value of $n_f$ and an uncertainty estimate for this average is obtained from the standard deviation. These results are also given in Tab.~\ref{tab:postdictiongluonium} and are shown in Fig.~\ref{fig:g4} as the points with error bars. The next step is the extraction of the coefficient of each $n_f$ power, $g_{4,1}^{(k)}$, which is done through a fit imposing the exact large-$n_f$ result, with the polynomial form
\beq
g_{4,1}(n_f)=\sum_{k=0}^3 g_{4,1}^{(k)} \, n_f^k + 0.0025809\,n_f^4,
\eeq
where $g_{4,1}^{(0)}$, $g_{4,1}^{(1)}$, $g_{4,1}^{(2)}$, and $g_{4,1}^{(3)}$ are the free parameters. The fit is not completely trivial because the points obtained from the averages of the PAs results are highly correlated and their covariance matrix turns out to be singular, since only three PAs are available in this case. Since the covariance matrix cannot be inverted, in order to perform a reliable fit, we resort to what we refer to as `diagonal fits'~\cite{Boito:2011qt,Boito:2022rad,Bruno:2022mfy}, where only the diagonal elements of the covariance matrix are kept in the modified $\chi^2$ function. In the error propagation, however, all correlations are taken into account through a linear error propagation, as described, for example, in App. A of Ref.~\cite{Boito:2011qt}. With a fit of this type we obtain
\begin{equation}
g_{4,1} (n_f) = (195 \pm 46) - (100 \pm 16) \, n_f + (10 \pm 4) \, n_f^2 - (0.1 \pm 0.5) \, n_f^3 + 0.0025809 \, n_f^4,
\label{eq:g4pred}
\end{equation}
which is also illustrated in Fig.~\ref{fig:g4} as the solid line. This estimate is in agreement with the true value provided in Eq.~(\ref{eq:nfdependenceex}), and the central values deviate by less than 25\% up to order $n_f^2$. Considering five active light quark flavors, we get $g_{4,1}(5) = -73 \pm 20$, where the correlations between the fit parameters $g_{4,1}^{(k)}$ are taken into account when propagating the uncertainties in the final result. As one can notice, this value agrees with the true one, $g_{4,1}(5) = -75.259$, differing by only 3\%. In
Tab.~\ref{tab:g4comp} we give the estimates for $g_{4,1}(n_f)$ for $n_f=1$ to $6$, together with the corresponding exact results. Our postdictions are consistent with the known results within the quoted uncertainties, with the relative deviation of the central values from the exact results remaining below~26\%.\footnote{Here, and throughout this paper, because of the cancellations between the contributions of successive powers of $n_f$, to obtain the exact results at specific $n_f$ values it is important to retain more digits than given in Eq.~(\ref{eq:g4pred}).}

\begin{table}[t]
\centering
\begin{tabular}{ccccccc}
\thickhline
& $g_{4,1}(1)$ & $g_{4,1}(2)$ & $g_{4,1}(3)$ & $g_{4,1}(4)$ & $g_{4,1}(5)$ & $g_{4,1}(6)$ \\ \hline
This work & $104 \pm 36$ & $33 \pm 31$ & $-20 \pm 29$ & $-55 \pm 25$ & $-73 \pm 20$ & $-74 \pm 17$ \\
Exact value & 118 & 41 & $-16$ & $-54$ & $-75$ & $-82$ \\ \thickhline
\end{tabular}
\caption{Predicted values of $g_{4,1}(n_f)$ of $\operatorname{Im} \Pi_{G^2}$ obtained using our method compared with the corresponding exact values.}
\label{tab:g4comp}
\end{table}

Continuing with the other series that are relevant for the $H\to gg$, the same procedure can be applied for the postdiction of the coefficient $d_{4,0}$ of the decoupling constant $\zeta_g^2$. This proves to be more challenging since the number of non-trivial coefficients available is smaller (only two) which significantly limits the number of approximants that can be built and their ability to forecast the subsequent coefficients. In this case, to mimic the situation we face in the prediction of $d_{5,0}$ we do not impose constraints from large-$n_f$ since these results are not known for higher orders. Only two approximants survive the consistency checks that we apply, namely, $P_1^1(u)$ and $\mathrm{Dlog}_1^1(a_s)$, and the respective results for $d_{4,0}(n_f)$ are shown in Tab.~\ref{tab:postdictiondecoupling} in App.~\ref{app:results}. After the fitting procedure we obtain
\begin{equation}
d_{4,0} (n_f) = (8 \pm 2) - (1.4 \pm 0.3) \, n_f + (0.06 \pm 0.01) \, n_f^2.
\label{eq:d4pred}
\end{equation}
The predictions of the first two terms, $d_{4,0}^{(0)}$ and $d_{4,0}^{(1)}$, agree with the true values, given in Eq.~(\ref{eq:nfdependencedecoupling}), within $2\sigma$. However, the estimated value for $d_{4,0}^{(2)}$ has the opposite sign of the exact result and differs by more than eight times the quoted uncertainty. This carries over to the value of the coefficient for a given $n_f$. For 5 light-quark flavors we obtain $d_{4,0}(5) = 2.6 \pm 0.5$, with a central value that has the opposite sign and lies about $6\sigma$ from the exact result, see Tab.~\ref{tab:d40comp}. These poor estimates, however, can be attributed to the fact that, for this perturbative series, only two approximants without defects can be used to postdict $d_{4,0}$ and the large-$n_f$ results are not imposed. As corroborated by the results we obtain for other series, for which more coefficients are available, we expect the forecast of $d_{5,0}$ to be improved with respect to this admittedly poor description of~$d_{4,0}$.

\begin{table}[t]
\centering
\begin{tabular}{ccccccc}
\thickhline
& $d_{4,0}(1)$ & $d_{4,0}(2)$ & $d_{4,0}(3)$ & $d_{4,0}(4)$ & $d_{4,0}(5)$ & $d_{4,0}(6)$ \\ \hline
This work & $7 \pm 1$ & $6 \pm 1$ & $4.4 \pm 0.9$ & $3.5 \pm 0.7$ & $2.6 \pm 0.5$ & $1.9 \pm 0.4$ \\
Exact value & 4.1 & 3.1 & 1.9 & 0.78 & $-0.43$ & $-1.7$ \\ \thickhline
\end{tabular}
\caption{Predicted values of $d_{4,0}(n_f)$ of $\zeta_g^2$ obtained using our method compared with the corresponding exact values.}
\label{tab:d40comp}
\end{table}

Postdicting, with the same method, the last known coefficient of the $\beta$ function, $\beta_5$ (in our conventions), we find
\begin{equation}
\beta_5 (n_f) = (909 \pm 352) - (148 \pm 58) \, n_f - (14 \pm 25) \, n_f^2 + (2 \pm 2) \, n_f^3 - 0.0035986 \, n_f^4.
\label{eq:beta5pred}
\end{equation}
Here the large-$n_f$ result was imposed since it is known to all orders and it can be used in the forecast of $\beta_6$. The approximants used to obtain this result are listed in Tab.~\ref{tab:postdictionbeta} in App.~\ref{app:results}.
The central values of the coefficients in our polynomial prediction show better agreement with the exact result at lower orders in $n_f$, with deviations below 60\% up to $\mathcal{O}(n_f)$. However, the increasing deviations at higher orders do not necessarily translate into poor predictions for $\beta_5(n_f)$ at fixed values of $n_f$.
The values of $\beta_5(n_f)$ obtained from our prediction in Eq.~(\ref{eq:beta5pred}) for different values of $n_f$ are presented in Tab.~\ref{tab:beta5comp}, together with the exact values and the corresponding predictions from Ref.~\cite{Ellis:1997sb}, which employs the method of the Asymptotic Pad\'e Approximants (APP). For smaller values of $n_f$, our central values show good agreement with the exact results, differing approximately by 4\%, 27\% and 60\% for $n_f=1,2,3$ respectively. Although the relative deviation increases for larger $n_f$, the exact values remain within $2\sigma$ of our predictions. In contrast, the results from Ref.~\cite{Ellis:1997sb} exhibit substantially larger deviations from the exact values, particularly for larger $n_f$.

\begin{table}[t]
\centering
\begin{tabular}{ccccccc}
\thickhline
& $\beta_5(1)$ & $\beta_5(2)$ & $\beta_5(3)$ & $\beta_5(4)$ & $\beta_5(5)$ & $\beta_5(6)$ \\ \hline
This work & $749 \pm 333$ & $577 \pm 286$ & $406 \pm 220$ & $252 \pm 156$ & $128 \pm 76$ & $49 \pm 30$ \\
Ref.~\cite{Ellis:1997sb} & 1096 & 787 & 557 & 406 & 331 & 332 \\
Exact value & 719 & 456 & 255 & 114 & 30 & 0.53 \\ \thickhline
\end{tabular}
\caption{Predicted values of $\beta_5(n_f)$ obtained using our method and the method reported by Ref.~\cite{Ellis:1997sb}, compared with the corresponding exact values.}
\label{tab:beta5comp}
\end{table}

Using the result for $d_{4,0}(n_f)$ given in Eq.~(\ref{eq:d4pred}), with the logarithms in $d_{n,k}$ resummed, together with the postdiction of $\beta_5$ given in Eq.~(\ref{eq:beta5pred}), we can determine the fourth-order coefficient of the Wilson coefficient $C_1$ using Eq.~(\ref{eq:c4}) given in App.~\ref{app:RGeq}. Taking into account the correlations within the sets of parameters $d_{4,0}^{(k)}$ and $\beta_5^{(k)}$, and assuming that the two sets are uncorrelated, we obtain for $\mu^2=m_t^2(\mu^2)$
\begin{equation}
c_4 (n_f) = -(336 \pm 168) + (227 \pm 132) \, n_f - (29 \pm 19) \, n_f^2 + (0.22 \pm 0.04) \, n_f^3.
\label{eq:c4pred}
\end{equation}
For $n_f = 5$, we find $c_4(5) = 89 \pm 174$, whose central value differs from the exact result by 146\%. Since this result is obtained directly from the decoupling relation, this relatively poor postdiction can be associated with the small number of non-trivial coefficients that can be used as input for $d_{4,0}$ (only two). Our estimates of $c_4$ for other physical values of $n_f$ are also consistent with the exact results within the corresponding uncertainties (except for $n_f=1$), although some of the central values exhibit significant deviations from the exact results, as shown in Tab.~\ref{tab:c4comp}. Therefore, this example should not be viewed as representative of the predictive power of the method in situations where a larger number of approximants is available. We show, however, that even in this case the postdiction of the coefficient $\gamma_4$ is very reasonable, since it is not very affected by $d_{4,0}$.

\begin{table}[t]
\centering
\begin{tabular}{ccccccc}
\thickhline
& $c_4(1)$ & $c_4(2)$ & $c_4(3)$ & $c_4(4)$ & $c_4(5)$ & $c_4(6)$ \\ \hline
This work & $-138 \pm 177$ & $2 \pm 210$ & $85 \pm 226$ & $114 \pm 213$ & $89 \pm 174$ & $11 \pm 128$ \\
Exact value & 251 & 193 & 138 & 86 & 36 & $-12$ \\ \thickhline
\end{tabular}
\caption{Predicted values of $c_4(n_f)$ of the Wilson coefficient $C_1$ at $\mu^2=m_t^2(\mu^2)$ obtained employing the postdiction of $d_{4,0}$ and $\beta_5$ given respectively in Eqs.~(\ref{eq:d4pred})~and~(\ref{eq:beta5pred}) compared with the corresponding exact values.}
\label{tab:c4comp}
\end{table}

With the postdictions of $g_{4,1}$, $d_{4,0}$ and $\beta_5$ given respectively in Eqs.~(\ref{eq:g4pred}),~(\ref{eq:d4pred})~and~(\ref{eq:beta5pred}), we can finally obtain the fourth-order coefficient of the decay rate of Higgs into gluons in the $\overline{\mathrm{MS}}$ scheme at $\mu^2 = m_H^2$ as
\begin{equation}
\gamma_4 (n_f) = (425 \pm 46) - (159 \pm 17) \, n_f + (13 \pm 4) \, n_f^2 - (0.1 \pm 0.5) \, n_f^3 + 0.0025809 \, n_f^4,
\label{eq:gamma4pred}
\end{equation}
where we take into account the correlations within the sets of parameters $d_{4,0}^{(k)}$, $\beta_5^{(k)}$ and $g_{4,1}^{(k)}$, while assuming that the different parameter sets are mutually uncorrelated. As per Eq.~(\ref{eq:gamma4-components}), the estimate of the Wilson coefficient does not play an important role in $\gamma_4$, as this quantity is mainly dominated by the coefficient of the gluonium correlator. In addition, the highest-order coefficient in $n_f$, as discussed in Sec.~\ref{sec:higgsdecaysqcd}, is the same as that of $\operatorname{Im} \Pi_{G^2}$. Considering $n_f=5$, we get
\[\gamma_4(5) = -68 \pm 20,\] which is consistent with the exact value within the quoted uncertainty. This agreement with the exact result is also observed for other values of $n_f$, as shown in Tab.~\ref{tab:gamma4comp}. This result is reassuring, since the predictions for the first unknown coefficient, benefiting from one more coefficient in each series, are expected to be better than the postdictions we obtained in this section.

\begin{table}[t]
\centering
\begin{tabular}{ccccccc}
\thickhline
& $\gamma_4(1)$ & $\gamma_4(2)$ & $\gamma_4(3)$ & $\gamma_4(4)$ & $\gamma_4(5)$ & $\gamma_4(6)$ \\ \hline
This work & $278 \pm 36$ & $156 \pm 32$ & $58 \pm 29$ & $-16 \pm 26$ & $-68 \pm 20$ & $-97 \pm 17$ \\
Exact value & 300 & 167 & 63 & $-15$ & $-71$ & $-106$ \\ \thickhline
\end{tabular}
\caption{Predicted values of $\gamma_4(n_f)$ obtained using our method compared with the corresponding exact values.}
\label{tab:gamma4comp}
\end{table}

\begin{figure}[t]
\centering
\includegraphics[width=0.5\textwidth]{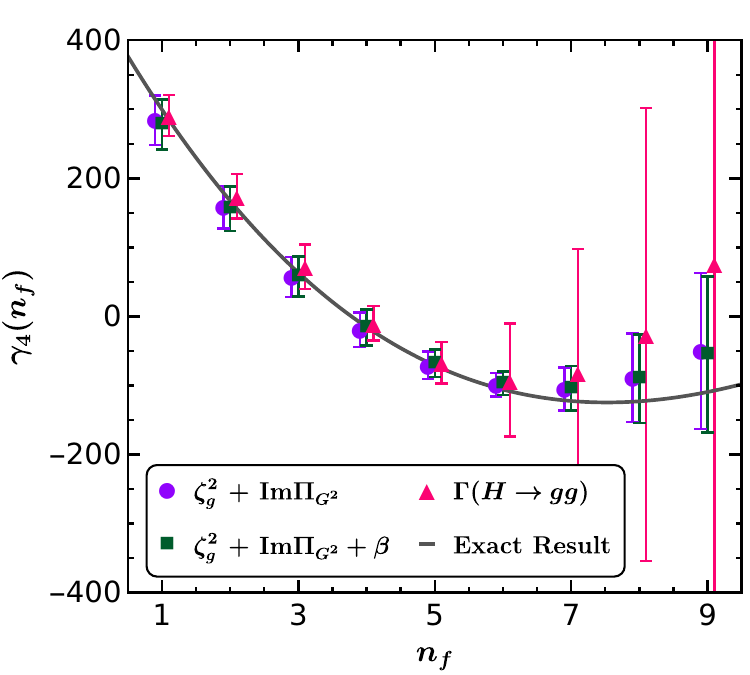}
\caption{Comparison of the predicted values of $\gamma_4(n_f)$ in the $\overline{\mathrm{MS}}$ scheme at $\mu=m_H$. Green squares show the result obtained by applying the method to each perturbative series independently, while pink triangles correspond to applying the approximants directly to $\Gamma(H\to gg)$. Purple dots show the result obtained using the known value of $\beta_5$, and the gray solid line represents the exact result. The coefficients are computed for integer values of $n_f$, but the data points are horizontally offset for clarity.}
\label{fig:compgamma4}
\end{figure}

Alternatively, the dependence of the coefficient $\gamma_4$ on the number of flavors could also have been obtained by applying the method directly to the perturbative series for the decay rate of Higgs into gluons. The disadvantage, in this case, is that the procedure would have to be repeated for each choice of scheme for the top-quark mass and for each value of the renormalization scale $\mu$. By estimating each part independently, as we did, we can instead obtain $\gamma_4$ for arbitrary values of $\mu$ and in any scheme.
For completeness, a comparison between these two possible strategies in the $\overline{\mathrm{MS}}$ scheme at $\mu^2 = m_H^2$ is shown in Fig.~\ref{fig:compgamma4}. The green squares represent the values of $\gamma_4(n_f)$ obtained by applying the method to each perturbative series independently, as discussed in this section, while the pink triangles correspond to the values obtained by applying the approximants directly to $\Gamma(H \to gg)$. It can be seen that the method explored in this section yields estimates that are closer to the exact result. Moreover, using the known result of the $\beta$-function coefficient $\beta_5$, instead of our estimate given in Eq.~(\ref{eq:beta5pred}), does not significantly affect the final result for $\gamma_4$, as shown by the purple dots in Fig.~\ref{fig:compgamma4}. This lack of sensitivity to the value of $\beta_{n+1}$ stems, partially, from the fact that the coefficient $c_n$ depends on $\beta_{n+1}(n_f)-\beta_{n+1}(n_f+1)$ [see Eqs.~(\ref{eq:c4}) and (\ref{eq:c5}) in App.~\ref{app:RGeq}]. Therefore, the constant term in Eq.~(\ref{eq:beta5pred}) cancels and the final uncertainty in this difference is somewhat reduced due to the correlations between the fit parameters.

The same procedure was also applied to the postdiction of the last known perturbative coefficient of the Higgs decay rate into a $b \bar{b}$ pair, $p_4$. In this case, we have two perturbative series at our disposal: the imaginary part of the scalar correlator, $\operatorname{Im} \Pi(s)$, which is directly related to $\Gamma(H \to b \bar{b})$, and the second derivative of the scalar correlator, $\Pi^{''}(s)$. However, in a previous work, Ref.~\cite{Boito:2021scm}, where we estimated the coefficient of order $\alpha_s^5$ of $\Gamma(H \to b \bar{b})$ based on the analysis of PAs and D-log PAs constructed considering only $n_f=5$, the results obtained with the approximants applied to $\Pi''(s)$ were more stable than those applied to $\operatorname{Im} \Pi(s)$. As discussed in Ref.~\cite{Boito:2021scm}, a possible explanation for this fact can be found in the large-$\beta_0$ approximation, where one sees that the Borel transform of $\Pi''$ is simpler, without the prefactor $\sin(\pi u)$ that appears in $B[\operatorname{Im}\Pi](u)$.
For this reason, the method employed in this work is applied only to $\Pi''(s)$. As mentioned earlier, the only independent coefficients are the $h_{n,1}$ which can be easily determined from the coefficients $k_n$ of $\Pi''(s)$, and through the RGE, we can then obtain the desired coefficient, $p_4$, of $ \operatorname{Im} \Pi(s)$.

Applying the consistency checks in this case we are left with seven approximants, which are all listed in Tab.~\ref{tab:postdictionhbb} in App.~\ref{app:results}. The method yields
\begin{equation}
p_4 (n_f) = (2 \pm 5) - (3 \pm 2) \, n_f + (0.2 \pm 0.2) \, n_f^2 - 0.000210041 \, n_f^3.
\label{eq:p4pred}
\end{equation}
All coefficients $p_4^{(k)}$ are consistent with their exact values, given in Eq.~(\ref{eq:hbbnfdependence}), within the estimated uncertainties. Taking into account the correlations between the fitted parameters, for $n_f=5$ we obtain $p_4(5)= -8.8 \pm 0.3$, whose central value differs from the exact result by only 4\%.

\begin{figure}[t]
\centering
\includegraphics[width=0.8\textwidth]{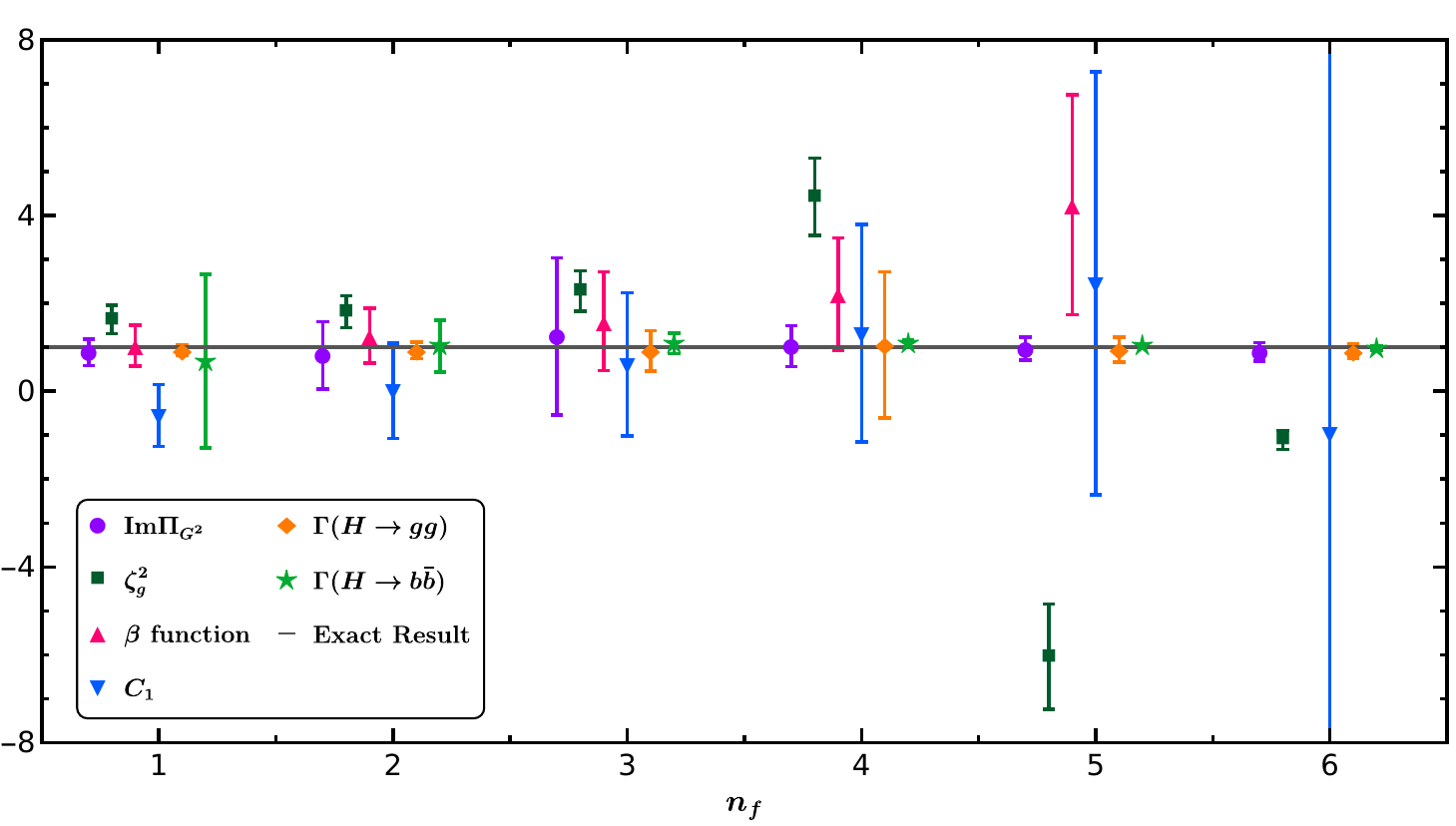}
\caption{Ratios of the predicted coefficients obtained from our method to their corresponding exact values for $n_f=1$ to 6, with the horizontal line indicating the exact result. Error bars represent the estimated uncertainties of the predictions. The results are computed for integer values of $n_f$ but are shown horizontally offset for clarity. The result for $\beta_5(n_f=6)$ lies outside the range of the figure.}
\label{fig:postdictions}
\end{figure}

In Fig.~\ref{fig:postdictions}, we summarize the postdictions discussed above by showing, for each perturbative series, the ratio of the predicted coefficient to its exact value for $n_f=1,\ldots,6$. The comparison illustrates both the predictive power of the procedure when sufficient information is available and its limitations when the available perturbative information is scarce. In particular, the postdiction of $\zeta_g^2$ is systematically displaced from the exact result since only two non-trivial coefficients can be used in this case, Eq.~(\ref{eq:nfdependencedecoupling}). This limitation propagates to the postdiction of the Wilson coefficient $C_1$ obtained from $\zeta_g^2$, whose central values can differ substantially from the exact result, although the relatively large uncertainties encompass the exact results for most $n_f$ values. On the other hand, the series for which more coefficients and, consequently, more consistent approximants are available generally lead to postdictions that are closer to the exact results and for which the exact values are well represented by the estimated uncertainties. This is particularly evident for $\operatorname{Im}\Pi_{G^2}$ and $\Gamma(H\to gg)$, as well as for the scalar correlator.

A final comment on the interpretation of our uncertainties is in order. Since the final uncertainties are obtained from a fit procedure, they inevitably inherit a statistical interpretation, and some of the true values of the coefficients will be outside a $1\sigma$ interval.

Having validated the procedure through these postdictions, we now turn to applications to the unknown higher-order coefficients of the series analyzed in this section.

\section{Results in QCD}\label{sec:results}
In this section, we present the results for the prediction of the first unknown perturbative QCD coefficient of the decoupling constant $\zeta_g^2$, of the imaginary part of the gluonium correlator, of the QCD $\beta$ function, of the Wilson coefficient $C_1$, and of the Higgs decay rate into two gluons. Results for the decay $H\to b\bar b$ will also be given. We employ the method introduced in the previous section, but now applied to the prediction of the first unknown coefficient of each series.

We start once again with the imaginary part of the gluonium correlator, $\operatorname{Im} \Pi_{G^2}$. After building the approximants to $\operatorname{Im} \Pi_{G^2}$, we apply the same criteria used in the previous section to veto approximants with Froissart doublets, with real poles very far from the origin, and/or with extraneous complex poles near the origin. The PAs and D-log PAs that pass all reliability tests discussed in Sec.~\ref{sec:method} are listed in Tab.~\ref{tab:predictiongluonium} in App.~\ref{app:results}. Three of the PAs are discarded, as indicated in Tab.~\ref{tab:predictiongluonium}, due to the values of $\Delta_{n_f}$, which are orders of magnitude larger than what we find for other PAs in the same table.

\begin{figure}[t]
\centering
\subfloat[]{\label{fig:g5}{\includegraphics[width=0.33\textwidth]{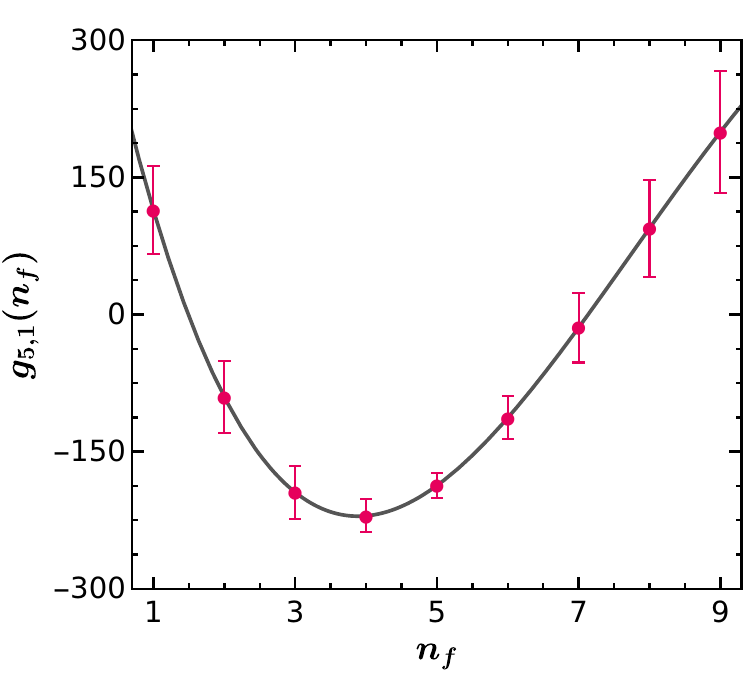}}} \hfill
\subfloat[]{\label{fig:d50}{\includegraphics[width=0.32\textwidth]{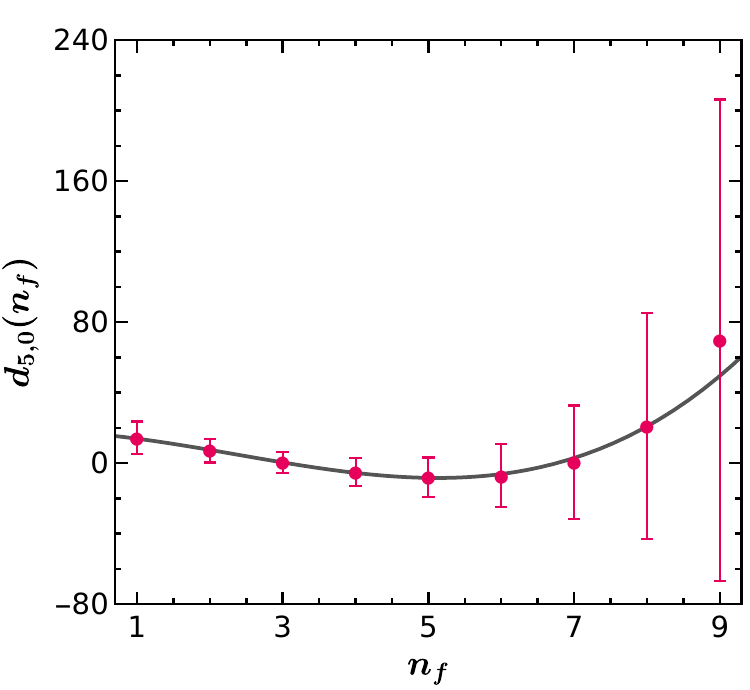}}} \hfill
\subfloat[]{\label{fig:beta6}{\includegraphics[width=0.33\textwidth]{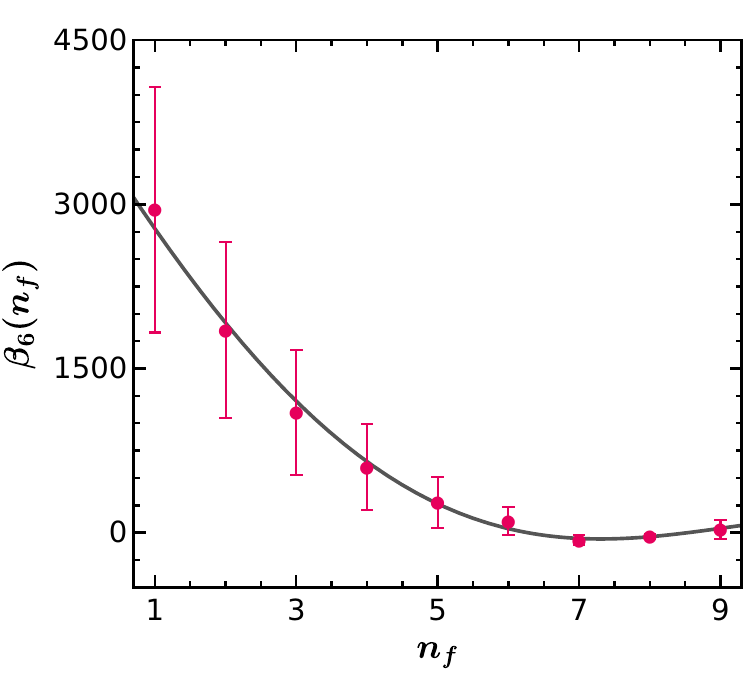}}}
\caption{Estimate of the first unknown coefficient of (a) $\operatorname{Im} \Pi_{G^2}$, (b) $\zeta_g^2$, (c) $\beta$ function obtained through the average of the predictions from different approximants. The polynomial fits for (a) $g_{5,1}(n_f)$ from Eq.~(\ref{eq:g5}), (b) $d_{5,0}(n_f)$ from Eq.~(\ref{eq:d50}), (c) $\beta_6(n_f)$ from Eq.~(\ref{eq:beta6}) are represented by the solid lines.}
\end{figure}

The prediction for each value of $n_f$ is then computed as before: the central value is the mean of the PAs and D-log PAs estimates and the uncertainty is the standard deviation of these predictions. Performing the same type of fit that we introduced in the previous section to these averaged predictions, but now with a polynomial of degree 5 in $n_f$, and imposing the large-$n_f$ result $g_{5,1}^{(5)} = -0.0002821$, we get
\begin{align}
g_{5,1}(n_f) = &(441 \pm 54) - (397 \pm 5) \, n_f + (74 \pm 2) \, n_f^2 \nonumber\\
&- (4.3 \pm 0.2) \, n_f^3 + (0.07 \pm 0.01) \, n_f^4 -0.0002821 \, n_f^5.
\label{eq:g5}
\end{align}
The outcome of this fit is compared with the predictions at each value of $n_f$ in Fig.~\ref{fig:g5}. Although the individual uncertainties of the coefficients $g_{5,1}^{(k)}$ are large, they are strongly correlated. Consequently, the propagated uncertainty for $g_{5,1}(n_f)$ can be substantially smaller than would be inferred from the individual parameter uncertainties, which can be seen in Fig.~\ref{fig:g5}. For example, for $n_f=5$, taking into account the correlations among fit parameters we obtain
\begin{equation}
g_{5,1}(5) = -187 \pm 14,
\end{equation}
where the resulting uncertainty of our estimate is 7.5\%, illustrating that the correlations among the fit parameters substantially affect the uncertainty of the polynomial evaluated at fixed $n_f$. Our estimate for $n_f=5$ quark flavors favors, therefore, a negative value for the fifth-order coefficient of $\operatorname{Im} \Pi_{G^2}$. The correlations among the fit parameters as well as the fit parameters with many more digits are given in App.~\ref{app:fits} --- we stress that, due to cancellations, it is important to use the parameters with more digits than given in Eq.~(\ref{eq:g5}) in order to obtain our final results.

We can now analyze the decoupling constant $\zeta_g^2$ in order to obtain an estimate for the unknown coefficient $d_{5,0}$. The approximants used are given in App.~\ref{app:results} in Tab.~\ref{tab:predictiondecoupling}. Our estimate for $d_{5,0}$ as a function of $n_f$ is
\begin{equation}
d_{5,0}(n_f) = (18 \pm 11) - (2 \pm 7) \, n_f - (2 \pm 3) \, n_f^2 + (0.3 \pm 0.4) \, n_f^3,
\label{eq:d50}
\end{equation}
which was obtained by fitting the data shown in Fig.~\ref{fig:d50}. In contrast to the fit for $g_{5,1}$, the highest-order coefficient in $n_f$, $d_{5,0}^{(3)}$, is not imposed, since its value is currently unknown. For $n_f=5$ active quark flavors, we find
\begin{equation}
d_{5,0}(5) = -8 \pm 11.
\end{equation}
Our estimate does not allow us to determine the sign of $d_{5,0}$ but, again, it favors a negative value. It is also worth noting that the magnitude of the central value is approximately 20 times larger than that of the last known coefficient, $d_{4,0}$. However, given the large uncertainty of our estimate, this difference should not be interpreted as evidence for a definite growth pattern of the perturbative coefficients.

The last quantity we need to consider before moving to the Wilson coefficient $C_1$ is the $\beta$-function. The approximants suitable to be used to estimate the coefficient $\beta_6$ are given in App.~\ref{app:results} in Tab.~\ref{tab:predictionbeta}. By performing a diagonal fit to the data exhibited in Fig.~\ref{fig:beta6}, one obtains the following $n_f$ polynomial dependence for $\beta_6$
\begin{align}
\beta_6(n_f) = &(3757 \pm 962) - (1025 \pm 211) \, n_f + (38 \pm 115) \, n_f^2 \nonumber \\
&+ (8 \pm 14) \, n_f^3 - (0.5 \pm 0.5) \, n_f^4 + 0.0017135 \, n_f^5,
\label{eq:beta6}
\end{align}
which is illustrated in Fig.~\ref{fig:beta6} by the solid line. Taking into account the correlations among the fit parameters $\beta_6^{(k)}$, we obtain the following estimate for the sixth-order coefficient of the $\beta$-function at $n_f=5$
\begin{equation}
\beta_6(5) = 264 \pm 226.
\label{eq:beta6nf5}
\end{equation}
Our final value of $\beta_6(5)$ has an uncertainty of about $85\%$. In the final value of $\gamma_5$ it enters through the combination $\beta_6(5)-\beta_6(6)$ that appears in $c_5$ [see Eq.~(\ref{eq:c5})], for which the uncertainty is somewhat reduced,
\beq
3(\beta_6(5) -\beta_6(6)) = 682 \pm 493.
\eeq
An estimate for $\beta_6(5)$ was also obtained in Ref.~\cite{Abbas:2022wnz} from Asymptotic Pad\'e Approximants (APP): $\beta_6^{\rm APP} = 109.6$ (in our conventions). No uncertainty is assigned to this latter estimate, but it is compatible with ours.

With the estimates of the unknown coefficients of the QCD $\beta$ function and the decoupling relation, we can obtain a prediction for the fifth-order coefficient of the Wilson coefficient $C_1$ through Eq.~(\ref{eq:c5}). The polynomial dependence of $c_5$ on the number of light-quark flavors using $\mu^2 = m_t^2(\mu^2)$, which resums the logarithms $L_t$, is
\begin{align}
c_5(n_f) = &-(873 \pm 767) + (574 \pm 740) \, n_f \nonumber \\
&+ (42 \pm 247) \, n_f^2 - (23 \pm 38) \, n_f^3 + (1 \pm 2) \, n_f^4,
\label{eq:c5poly}
\end{align}
where, in propagating the uncertainty, we accounted for the correlations among the parameters $d_{5,0}^{(k)}$ and $\beta_6^{(k)}$, while assuming no correlations between the two sets of parameters. The highest-order coefficient in $n_f$, $c_5^{(4)}$, receives a contribution from $d_{5,0}^{(3)}$ and inherits the associated uncertainty, as can be seen in Eq.~(\ref{eq:c5}) in App.~\ref{app:RGeq}. Considering 5 active light-quark flavors, we obtain
\begin{equation}
c_5(5) = 985 \pm 677.
\end{equation}
Once again, although the individual uncertainties of the fit parameters $c_5^{(k)}$ are very large, their correlations substantially reduce the propagated uncertainty when evaluating $c_5$ at a fixed value of $n_f$. For $n_f=5$, our estimate has a relative uncertainty of approximately 69\%. Moreover, our result favors a positive value of $c_5$.

Finally, we estimate the fifth-order coefficient of the Higgs decay rate into gluons, $\gamma_5$. Employing our estimates for $g_{5,1}(n_f)$, $d_{5,0}(n_f)$ and $\beta_6(n_f)$, given respectively in Eqs.~(\ref{eq:g5}),~(\ref{eq:d50})~and~(\ref{eq:beta6}), we obtain the following $n_f$ dependence for $\gamma_5$ in the $\overline{\mathrm{MS}}$ scheme at $\mu^2=m_H^2$
\begin{align}
\gamma_5 (n_f) = &(1201 \pm 54) - (719 \pm 7) \, n_f + (108 \pm 3) \, n_f^2 \nonumber \\
&- (5.5 \pm 0.3) \, n_f^3 + (0.09 \pm 0.02) \, n_f^4 - 0.0002821 \, n_f^5.
\label{eq:gamma5nf}
\end{align}
As discussed in Sec.~\ref{sec:higgsdecaysqcd}, the highest-order coefficient in $n_f$, $\gamma_5^{(5)}$, is exactly the same as the corresponding coefficient from the imaginary part of the gluonium correlator, $g_{5,1}^{(5)}$. Once again, the relatively large uncertainties of the fit parameters $\gamma_5^{(k)}$ do not result in comparably large uncertainties when evaluating the coefficient at a fixed value of $n_f$. Considering 5 light-quark flavors, we obtain
\begin{equation}
\gamma_5(5) = -312 \pm 15,
\label{eq:gamma5}
\end{equation}
where the correlations within the set of parameters $d_{5,0}^{(k)}$, $\beta_6^{(k)}$, and $g_{5,1}^{(k)}$ are taken into account and we used $\mu^2=m_H^2$. Following the same procedure as in Sec.~\ref{sec:method}, the correlations between different sets are not considered. This result can be compared with the estimate of Ref.~\cite{Abbas:2022wnz} which can be translated, in our conventions, to the following estimate of $\gamma_5(5)$ in the $\overline{\mathrm{MS}}$ scheme with $\mu^2=m_H^2$: $\gamma_5^{\rm APP}=-282\pm 54$. Our result is compatible with this value.

\begin{figure}[t]
\centering
\includegraphics[width=0.5\textwidth]{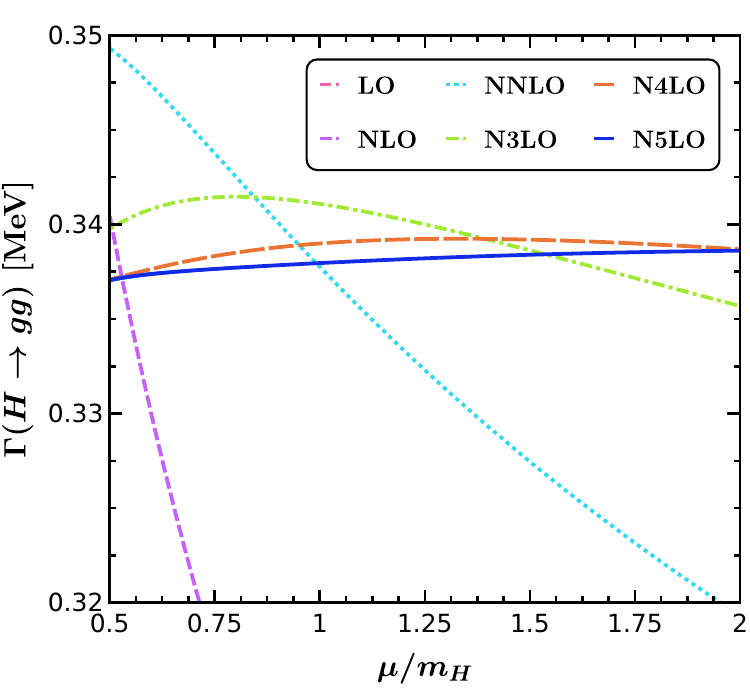}
\caption{Scale dependence of the decay width of $H \to gg$ computed in the $\overline{\mathrm{MS}}$ scheme in the interval $m_H/2 \leq \mu \leq 2m_H$.}
\label{fig:muevolutionHgg}
\end{figure}

Using these estimates, we can assess the impact of the inclusion of $\gamma_5$ on the contribution of $C_1$ to the Higgs boson decay width into two gluons. The residual renormalization-scale dependence of the perturbative expansion of $\Gamma(H \to gg)$ arises from the running of $\alpha_s$, from the explicit scale dependence in the logarithmic terms, and from the running of the top-quark mass in these logarithms. The logarithms have two origins. The first type is associated with the dependence of $\operatorname{Im}\Pi_{G^2}$ on the external momentum and the renormalization scale, and is given by $L_H= \ln{(m_H^2/\mu^2)}$. The second type stems from the decoupling relation and is described by $L_t = \ln{(\mu^2 / m_t^2(\mu^2))}$. Consequently, the coefficients $\gamma_n$, as indicated in Eq.~(\ref{eq:decayrateHgg}), acquire a dependence on these logarithms. The coefficients of the logarithms are fixed by RGEs once the non-logarithmic coefficients are known. Choosing $\mu^2 = s = m_H^2$ resums the logarithms $L_H$ and we recover the numerical results in Eq.~(\ref{eq:hggnfdependence}), where the remaining dependence on $L_t$ is evaluated with $\mu=m_H$. We evaluate this contribution to the decay rate as a function of the renormalization scale, varying $\mu$ in the range $m_H/2 \leq \mu \leq 2m_H$. Our result is illustrated in Fig.~\ref{fig:muevolutionHgg} for the $\overline{\mathrm{MS}}$ mass scheme, using our prediction for $\gamma_5$, Eq.~(\ref{eq:gamma5}), with the $\mu$ dependence reconstructed with the RG. The uncertainty shown for the N5LO result is solely due to the error in the predicted coefficient $\gamma_5$. The decay rate at N4LO still exhibits a sizable dependence on the renormalization scale. In contrast, the inclusion of our N5LO correction reduces the residual renormalization-scale dependence, resulting in a more stable prediction.

Considering the contribution from $C_1$ in Eq.~(\ref{eq:leffhgg}), the decay width of the Higgs boson into two gluons up to N5LO is
\begin{align}
\Gamma(H \to gg) = 0.33796 &\pm (0.00627)_{\alpha_s} \pm (0.00089)_{m_H} \nonumber \\
&\pm (0.00078)_\mu \pm (0.00006)_{m_t} \pm (0.00005)_{\gamma_5} \,\,\, \mathrm{MeV},
\end{align}
where we used the following values: $\alpha_s (m_Z) = 0.1180 \pm 0.0009$, $m_H = (125.13 \pm 0.11)$~GeV, $m_t(m_t^2) = 162.5^{+2.1}_{-1.5}$~GeV and $\gamma_5 = -312 \pm 15$.
The uncertainty marked with ``$\gamma_5$'' represents the error due to our predicted coefficient $\gamma_5$. The uncertainty corresponding to the renormalization scale was computed as half of the maximum difference of the decay rate when $\mu$ is varied in the range $m_H/2 \leq \mu \leq 2m_H$. The inclusion of our estimate for $\gamma_5$ reduces the scale-variation uncertainty by a factor of almost 1.5 when compared with the result truncated at $\gamma_4$. The uncertainty of the $H \to gg$ decay rate is highly dominated by the error associated with the value of the strong coupling constant, which gives an uncertainty about 5 times larger than all other sources of uncertainty combined in quadrature.
Our final result for the contribution of $C_1$ to the Higgs decay rate into two gluons is consistent with other estimates in the literature~\cite{Zeng:2018jzf,Zeng:2020lwi,Gao:2021wjn,Yan:2022foz,AlamKhan:2023dms}. The uncertainty from $\alpha_s$ also agrees with other works~\cite{Gao:2021wjn,AlamKhan:2023dms}.

We can now apply our method to update our previous estimate for the first unknown coefficient of the Higgs decay rate into bottom quarks, given in Ref.~\cite{Boito:2021scm}. This coefficient was previously estimated using a similar approach based on Pad\'e and D-log Pad\'e approximants, in a dedicated analysis that used large-$\beta_0$ results to guide the approximation at higher orders, but without exploring the known polynomial dependence of the coefficients on $n_f$ or imposing the large-$n_f$ results on the predicted coefficients.

\begin{figure}[t]
\centering \includegraphics[width=0.455\textwidth]{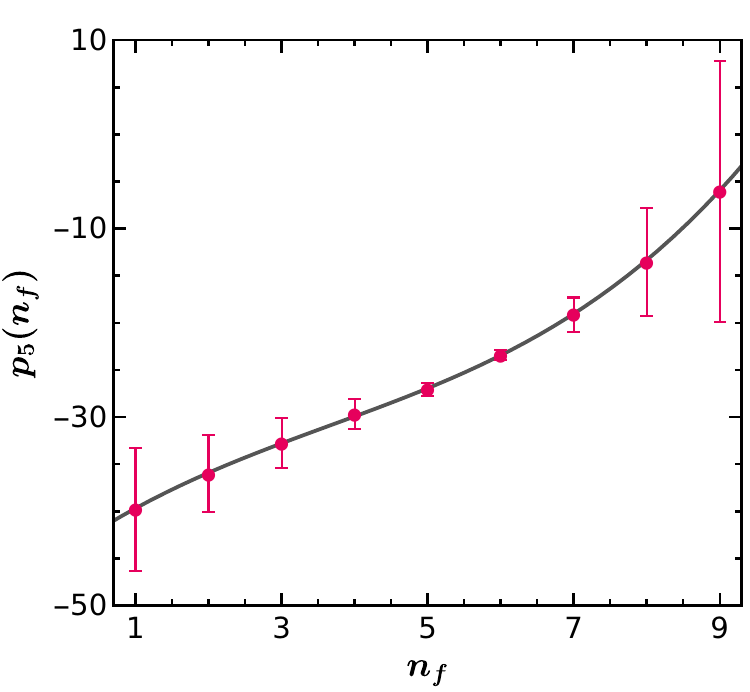}
\caption{Estimate of the coefficient $p_5(n_f)$ of the Higgs decay rate into $b \bar{b}$, obtained by averaging the predictions from different approximants, as a function of the number of flavors. The polynomial fit given by Eq.~(\ref{eq:c5Hbb}) is represented by the solid line.}
\label{fig:p5Hbb}
\end{figure}

We first consider the perturbative expansion of $\Gamma(H \rightarrow b \bar{b})$. We apply PAs and D-log PAs to the second derivative of $\Pi(s)$, given in Eq.~(\ref{eq:secderscalarcorrelator}). As mentioned earlier, the only independent coefficients are the $h_{n,1}$ which can be determined from the coefficients $k_n$ of $\Pi''(s)$. The desired coefficient $p_5$ of $ \operatorname{Im} \Pi(s)$ can then be obtained using the RG.
The resulting estimate for $p_5(n_f)$, obtained from the fit and shown by the solid line in Fig.~\ref{fig:p5Hbb}, is
\begin{equation}
p_5 (n_f) = -(45 \pm 12) + (5 \pm 8) \, n_f - (0.7 \pm 1.9) \, n_f^2 + (0.06 \pm 0.14) \, n_f^3 + 7.09 \times 10^{-6} \, n_f^4.
\label{eq:c5Hbb}
\end{equation}
The approximants employed to obtain this result are listed in Tab.~\ref{tab:predictionhbb} in App.~\ref{app:results}. As observed for other quantities, the uncertainties of $p_5^{(k)}$ are relatively large, but the strong correlations among these parameters substantially reduce the uncertainty of $p_5$ at fixed $n_f$. For $n_f=5$, we obtain
\begin{equation}
p_5(5) = -27.0 \pm 0.5,\label{eq:p5val}
\end{equation}
corresponding to a relative uncertainty of 1.9\%. Our previous result, reported in Ref.~\cite{Boito:2021scm}, was $p_5(5) = -23 \pm 5$, which is fully compatible with our new estimate.\footnote{Note that in Ref.~\cite{Boito:2021scm}, the corresponding coefficient (denoted $c_5$ there) is given for the perturbative series in $a_s$ of $\operatorname{Im} \Pi(s)$, rather than in $\alpha_s$ as used in this work. Thus, the result of Ref.~\cite{Boito:2021scm} must be divided by $\pi^5$ before comparison with Eq.~(\ref{eq:p5val}).}

\section{Conclusions}\label{sec:conclusions}

In this work we have obtained model-independent estimates for the first unknown perturbative QCD coefficients governing the hadronic decays of the Higgs boson using Pad\'e and D-log Pad\'e approximants built both from the perturbative series in $\alpha_{s}$ and from their Borel transforms.
Here we have obtained the coefficients as polynomials in powers of $n_f$ using the large-$n_f$ result as input, when available, in the spirit of what was done already in the past~\cite{Ellis:1996zn,Ellis:1997sb}. The large-$n_f$ results were also used as an additional test for the quality of the approximation obtained from individual approximants, and we have vetoed approximants that present important discrepancies with respect to the large-$n_f$ results. Since large-$n_f$ results are exact, our results remain model independent. In contrast, in previous analyses that some of us have co-authored~\cite{Boito:2018rwt, Boito:2021scm}, we have made use of large-$\beta_0$ results to guide the approximation which, albeit defensible, departs somewhat from strict model independence.

We validated the method through the postdictions of the last known coefficients of $\mathrm{Im}\,\Pi_{G^2}$, $\zeta_g^2$, the QCD $\beta$ function, and the imaginary part of the massless scalar correlator $\mathrm{Im}\,\Pi(s)$, all of which reproduce the exact results within the uncertainties with the exception of $\zeta_g^2$, for which only two non-trivial coefficients can be used to build the approximants, degrading the result for the postdiction of $d_{4,0}$ and, consequently, of $c_{4}$. Nevertheless, the postdiction of $\gamma_4$ is still successful, since it is dominated by $g_{4,1}$, giving confidence in applying the same procedure to the prediction of unknown coefficients. The decay $H\to gg$ relies on two ingredients: the absorptive part of the gluonium correlator, $\mathrm{Im}\,\Pi_{G^2}$, and the Wilson coefficient $C_1$.
The latter is obtained using a low-energy theorem in terms of derivatives of $\zeta_g^2$. In the reconstruction of its logarithms using the RG, higher-order coefficients of the $\beta$ function are also required. Furthermore, we have shown that the large-$n_f$ results for $H\to gg$ can already be obtained from results available in the literature, since the highest power of $n_f$ in the coefficients $\gamma_n$ is that of $g_{n,1}$, for which the large-$n_f$ results are known to all orders~\cite{Jamin:2012nn}.

Our main results concern the Higgs boson decay to gluons. For five active flavors, we obtain the fifth-order coefficients of the gluonium correlator, $g_{5,1}(5)=-187\pm 14$, and of the decoupling constant, $d_{5,0}(5)=-8\pm11$, together with the six-loop $\beta$-function coefficient, $\beta_6(5)=264\pm 226$. Combining them through the low-energy theorem and the associated RGE, these yield the first unknown coefficient of the Wilson coefficient, and, finally, of the $C_1$ contribution to the decay rate. For $n_f=5$ and using $\mu^2=m_t^2(\mu^2)$ in the $\overline {\rm MS}$ scheme we find
$c_5(5)=985\pm677$, and, for the decay rate, $\gamma_5(5)=-312\pm15$, where we used $\mu^2 = m_H^2$. This result is dominated by $g_{5,1}$ which could be determined with a smaller error than $\beta_6$ and $d_{5,0}$.

Including this N5LO estimate in the perturbative expansion of $\Gamma(H\to gg)$ reduces the residual renormalization-scale dependence of the decay rate: the associated uncertainty is smaller by a factor of about $1.5$ with respect to the N4LO result, when $\mu$ is varied in the range $m_H/2\le\mu\le 2m_H$. Our final result for the decay width implied by the Lagrangian of Eq.~(\ref{eq:leffhgg}), including the predicted coefficient $\gamma_5$, reads
\begin{align}
\Gamma(H \to gg) = 0.33796 &\pm (0.00627)_{\alpha_s} \pm (0.00089)_{m_H} \nonumber \\
&\pm (0.00078)_\mu \pm (0.00006)_{m_t} \pm (0.00005)_{\gamma_5} \,\,\, \mathrm{MeV},\label{eq:Gamma-Hgg}
\end{align}
which is consistent with other estimates in the literature~\cite{Zeng:2018jzf,Zeng:2020lwi,Gao:2021wjn,Yan:2022foz,AlamKhan:2023dms}. The strong coupling constant $\alpha_s$ is the dominant source of uncertainty, followed by the Higgs boson mass. The truncation uncertainty associated with $\gamma_5$ is subdominant, at present, and comparable with the top-mass uncertainty. However, with the forecast $0.1\%$ precision for $\alpha_s(m_Z)$ and the $4$~MeV uncertainty in $m_H$ that should be attained at the planned FCC-ee~\cite{FCC:2025lpp}, the uncertainty associated with the truncation of the QCD perturbative series, as measured by the residual scale variation, would become as large as the parametric uncertainties.

Applying the same procedure to $H\to b\bar b$, we determine the polynomial dependence on $n_{f}$ and update the estimate of Ref.~\cite{Boito:2021scm} for the first unknown coefficient of $\mathrm{Im}\,\Pi(s)$, finding $p_5(5)=-27.0 \pm 0.5$, in excellent agreement with the earlier result~\cite{Boito:2021scm}.

We conclude that this new strategy, combining Padé and D-log Padé approximants with constraints from exact large-$n_f$ results, provides reliable and model-independent estimates of the unknown higher-order corrections for the series we have discussed when sufficient information is available (at least three non-trivial coefficients of the perturbative series).
Given the increasing difficulty of exact multi-loop computations, we expect this type of approach to remain a useful tool for estimating missing higher-order corrections and assessing the associated truncation and scale uncertainties in advance of future exact calculations.

\section*{Acknowledgements}
DB’s work was supported
by the S\~ao Paulo Research Foundation (FAPESP) grant No. 2021/06756-6 and by CNPq
grant No. 303553/2025-1. CYL's work was supported by the S\~{a}o Paulo Research Foundation (FAPESP) under Ph.D.\ Grant No.~2020/15532-1 and postdoctoral Grant No.~2026/03498-0. GAN’s work was supported by the S\~{a}o Paulo Research Foundation (FAPESP) under M.Sc.\ Grant No.~2022/01861-9 and Ph.D.\ Grant No.~2024/01465-1.

\appendix

\section{Gluonium correlator in the large-\texorpdfstring{\boldmath{$\beta_0$}}{beta0} limit}\label{app:impig2beta0}

For completeness, we reproduce here the large-$\beta_0$ results for the gluonium correlator~\cite{Jamin:2012nn} from which we extract the coefficient of the highest power of $n_f$ of $g_{n,1}$ and, consequently, of $\gamma_{n}$.
The result for $\Pi_{G^{2}}$ in the large-$\beta_0$ limit can be written as~\cite{Jamin:2012nn}
\begin{align}
\Pi_{G^2}^{\text{large-}\beta_0}(s) &= \frac{(N_c^2-1)}{2\pi^2}\, s^2\,(3-2\varepsilon)\left(\frac{\mu^2 e^{\gamma_E}}{-s}\right)^{\varepsilon}
\sum_{m=0}^{\infty}\sum_{n=0}^{\infty}\left(-\frac{\beta_1 a_s}{2\hat{\varepsilon}}\right)^{m+n} \nonumber\\
&\sum_{k=0}^{m}\sum_{l=0}^{n}\binom{m}{k}\binom{n}{l}
\left[-6\,B(2-\varepsilon,2-\varepsilon)\,\Gamma(1+\varepsilon)\left(\frac{\mu^2 e^{\gamma_E}}{-s}\right)^{\varepsilon}\right]^{k+l} \nonumber\\
&\Big[\,2-(3+k+l)\varepsilon+(k+1)(l+1)\varepsilon^2\,\Big]\,G_{kl}(\varepsilon)\,,
\label{eq:cggb0}
\end{align}
where $1/\hat{\varepsilon} \equiv 1/\varepsilon - \gamma_E + \ln(4\pi)$, $B(x,y)$ is the Euler beta function, and the function $G_{kl}(\varepsilon)$ is found to be
\begin{equation}
G_{kl}(\varepsilon) = \frac{\Gamma[2-(k+1)\varepsilon]\,\Gamma[2-(l+1)\varepsilon]}
{\Gamma[1+k\,\varepsilon]\,\Gamma[1+l\,\varepsilon]\,\Gamma[4-(k+l+2)\varepsilon]}\,
\Gamma[-2+(k+l+1)\varepsilon]\,.
\label{eq:Gkl}
\end{equation}
Here the indices $m$ and $n$ count the bubbles inserted in each of the two gluon propagators, whereas the binomial sums over $k$ and $l$ implement the subtraction of the subdivergences, i.e. the renormalization of the coupling, which is thus already built into Eq.~(\ref{eq:cggb0}).

Therefore, expanding Eq.~\eqref{eq:cggb0} order by order in $a_s$, setting $\mu^2=s$, and taking the imaginary part of the finite term (with $s\to s+i0$), one obtains the large-$\beta_0$ result for $\mathrm{Im}\,\Pi_{G^2}$ to all orders. Collecting, order by order in $\alpha_{s}$, the term of highest degree in $n_f$, we write
\begin{equation}
g^{(n)}_{n,1}=\left(-\frac{1}{3\pi}\right)^{\!n} \kappa_n\,,
\label{eq:gnmaxkappa}
\end{equation}
where the $\kappa_n$ are pure numbers. With this normalization, naive
non-Abelianization, $n_f\to-3\beta_1$, gives simply
$g^{[L\beta_0]}_{n,1}=\kappa_n\,(\beta_1/\pi)^n$. For the first five orders we find
\begin{align}
\kappa_1 &= \frac{7}{2}\,, \nonumber\\
\kappa_2 &= \frac{127}{12}-\frac{\pi^2}{4}\,, \nonumber\\
\kappa_3 &= \frac{7127}{216}-\frac{7\pi^2}{4}-\zeta_3\,, \nonumber\\
\kappa_4 &= \frac{24157}{216}-\frac{635\pi^2}{72}+\frac{\pi^4}{16}
-\frac{35}{4}\,\zeta_3\,, \nonumber\\
\kappa_5 &= \frac{2175035}{5184}-\frac{35635\pi^2}{864}+\frac{21\pi^4}{32}
-\left(\frac{1279}{24}-\frac{5\pi^2}{4}\right)\zeta_3-6\,\zeta_5\,.
\label{eq:kappan}
\end{align}
Numerically, Eqs.~\eqref{eq:gnmaxkappa} and \eqref{eq:kappan} give
\begin{align}
g_{1,1}^{(1)} &= -0.37136\,, &
g_{2,1}^{(2)} &= \phantom{-}0.091368\,, &
g_{3,1}^{(3)} &= -0.017346\,, \nonumber\\
g_{4,1}^{(4)} &= \phantom{-}0.0025809\,, &
g_{5,1}^{(5)} &= -0.0002821\,; &
\label{eq:gnmaxb0}
\end{align}
where the first four monomials reproduce exactly those of Eq.~(\ref{eq:imgluoniumnfdependence}), obtained from the full QCD calculation, while $g_{5,1}^{(5)}$ is the leading-$n_{f}$ monomial of the first unknown coefficient, and is used as a constraint in the fit.

\section{Renormalization group equation for \texorpdfstring{\boldmath{$\zeta_g^2$}}{zeta}}
\label{app:RGeq}
Employing the RGE for $\zeta_g^2$, Eq.~(\ref{eq:zetagRGE}), it is possible to determine the coefficients $d_{n,k>0}$ of $\zeta_g^2$ in terms of the independent coefficients $d_{n,0}$ and the coefficients of the QCD $\beta$- and $\gamma_m$-functions. Our conventions for the $\beta$ function are given in Eq.~(\ref{eq:betafunc-def}) while for $\gamma_m$ we use
\beq
-\frac{\mu}{m_q} \frac{dm_q}{d\mu}
= \gamma_m(a_s)= \gamma_{m,1}\,a_s+\gamma_{m,2}\,a_s^2+\cdots,
\eeq
which leads to $\gamma_{m,1}=3 \, C_F/2$ and $\gamma_{m,2}=101/12-(5/18)n_f$. The coefficients $\gamma_{m,i}^\prime$ of Eq.~(\ref{eq:zetagRGE}) are obtained using $n_f+1$ flavors. After computing the logarithmic terms of the decoupling relation for higher orders, one can derive the perturbative coefficients $c_n$ of the Wilson coefficient $C_1$ in terms of $d_{n,k}$ through the low-energy theorem, Eq.~(\ref{eq:LET}). The fourth-order coefficient, $c_4$, as a function of $d_{4,0}$ and $\beta_5$ (which are both exactly known at present) is
\begin{multline}
c_4(n_f) = -\frac{88588363}{248832} - \frac{1974085}{18432} \,\zeta_3 + \frac{475}{9}\,\zeta_5 - \frac{31\pi^4}{432} -\frac{49187545}{55296} \, L_t + \frac{9364157}{12288} \,\zeta_3 \,L_t \\
+ \frac{8401}{384} \, L_t^2 + \frac{2299}{256} \, L_t^3 + \left( \frac{3221053}{55296} + \frac{3843215}{110592} \,\zeta_3 -\frac{575}{36} \,\zeta_5 - \frac{29\pi^4}{432} \right. \\
\left. + \frac{5160073}{41472} \,L_t - \frac{32257}{288} \, \zeta_3 \,L_t + \frac{4139}{2304} \,L_t^2 +\frac{275}{192} \,L_t^3 \right) \, n_f - \left( \frac{3503}{62208} + \frac{3}{4} \, \zeta_3 \right. \\
\left. - \frac{\pi^4}{216} + \frac{21139}{3456} \, L_t - \frac{28297}{9216} \, \zeta_3 \, L_t + \frac{1267}{13824} \, L_t^2 + \frac{157}{576} \,L_t^3 \right) \, n_f^2 + \left( \frac{83}{7776} - \frac{1}{54} \, \zeta_3 \right. \\
\left. + \frac{481}{5184} \,L_t - \frac{77}{6912} \, L_t^2 + \frac{1}{108} \, L_t^3 \right) \, n_f^3 - 3 \left(\frac{31}{2} - n_f \right) d_{4,0} +3\,(\beta_5-\beta_5^{\prime}) ,
\label{eq:c4}
\end{multline}
where $L_t = \ln (\mu^2/ m_t^2(\mu^2))$. The perturbative expansion here is in terms of the number of light flavors, so $n_f=5$ in our concrete application. In $L_t$, however, the running mass is in the full theory, with $n_f+1$ flavors.

For the fifth-order coefficient, $c_5$, the expression in terms of the unknown coefficients $d_{5,0}$, of $\zeta_g^2$, and $\beta_6$ is
\begin{multline}
c_5(n_f) = - \frac{20074434071}{27993600} - \frac{575469367}{55987200} \, \pi^4 - \frac{14725}{40824} \, \pi^6 + \frac{9318467}{1166400} \, \pi^4 \, \ln 2 \\
- \frac{3031309}{46656} \, \pi^2 \, \ln^2 2 + \frac{340853}{29160} \, \pi^2 \, \ln^3 2 + \frac{3031309}{46656} \, \ln^4 2 - \frac{340853}{48600} \, \ln^5 2 \\
+\frac{3031309}{1944} \, \operatorname{Li}_4\left(\frac12\right) + \frac{340853}{405} \, \operatorname{Li}_5\left(\frac12\right) - \frac{54457594489}{24883200} \, \zeta_3 - \frac{38015}{216} \, \zeta_3^2 \\
+ \frac{418544881}{311040} \, \zeta_5 -\frac{582715}{864} \, \zeta_7 - \frac{2575039948741}{130636800} \, L_t + \frac{51916497413}{261273600} \, \pi^4 \, L_t \\
- \frac{3177597247}{21772800} \, \pi^4 \, \ln 2 \, L_t + \frac{1033676369}{870912} \, \pi^2 \, \ln^2{2} \, L_t -\frac{116230873}{544320} \, \pi^2 \, \ln^3{2} \, L_t \\
- \frac{1033676369}{870912}\, \ln^4 2 \, L_t + \frac{116230873}{907200} \, \ln^5 2 \, L_t - \frac{1033676369}{36288} \, \operatorname{Li}_4\left(\frac12\right) \, L_t \\
- \frac{116230873}{7560} \, \operatorname{Li}_5\left(\frac12\right) \, L_t + \frac{139463800663}{9676800} \, \zeta_3 \, L_t + \frac{11137168559}{967680} \, \zeta_5 \, L_t \\
- \frac{544948211}{110592} \, L_t^2 + \frac{103005727}{24576} \, \zeta_3 \, L_t^2 + \frac{521083}{6912} \, L_t^3 +\frac{25289}{1024} \, L_t^4 \\
+ \left( \frac{19354335209}{37324800} - \frac{114666277}{27993600} \, \pi^4 + \frac{21625}{163296} \, \pi^6 + \frac{9318467}{9331200} \, \pi^4 \, \ln 2 \right. \\
-\frac{3088849}{373248} \, \pi^2 \, \ln^2 2 + \frac{340853}{233280} \, \pi^2 \, \ln^3 2 + \frac{3088849}{373248} \, \ln^4 2 -\frac{340853}{388800} \, \ln^5 2 \\
+ \frac{3088849}{15552} \, \operatorname{Li}_4\left(\frac12\right) + \frac{340853}{3240} \, \operatorname{Li}_5\left(\frac12\right) + \frac{14795064067}{24883200} \, \zeta_3 + \frac{4195}{432} \, \zeta_3^2 \\
- \frac{733202047}{1244160} \, \zeta_5 + \frac{56875}{432} \, \zeta_7
+ \frac{5301153383137}{1567641600} \, L_t - \frac{18084086501}{3135283200} \, \pi^4 \, L_t \\
+ \frac{9318467}{510300} \, \pi^4 \, \ln 2 \, L_t - \frac{761299249}{5225472} \, \pi^2 \, \ln^2 2 \, L_t + \frac{681706}{25515} \, \pi^2 \, \ln^3 2 \, L_t \\
+ \frac{761299249}{5225472} \, \ln^4 2 \, L_t - \frac{681706}{42525} \, \ln^5 2 \, L_t + \frac{761299249}{217728} \, \operatorname{Li}_4\left(\frac12\right) \, L_t \\
+ \frac{5453648}{2835} \, \operatorname{Li}_5\left(\frac12\right) \, L_t - \frac{490077699791}{116121600} \, \zeta_3 \, L_t - \frac{30129649}{22680} \, \zeta_5 \, L_t \\
\left. + \frac{161460679}{165888} \, L_t^2 - \frac{32073085}{36864} \, \zeta_3 \, L_t^2 + \frac{269}{10368} \, L_t^3 + \frac{3751}{1536} \, L_t^4 \right) \, n_f \\
+ \left( -\frac{3812693}{279936} + \frac{897029}{29859840} \, \pi^4 -\frac{575}{81648} \, \pi^6 - \frac{4795}{248832} \, \pi^2 \, \ln^2 2 \right. \\
+ \frac{4795}{248832} \, \ln^4 2 + \frac{4795}{10368} \, \operatorname{Li}_4\left(\frac12\right) - \frac{341945}{24576} \, \zeta_3 - \frac{725}{216} \, \zeta_3^2 + \frac{103405}{10368} \, \zeta_5 \\
- \frac{16440127651}{97977600} \, L_t - \frac{315918647}{195955200} \, \pi^4 \, L_t - \frac{9318467}{16329600} \, \pi^4 \, \ln 2 \, L_t \\
+ \frac{2801149}{653184} \, \pi^2 \, \ln^2 2 \, L_t - \frac{340853}{408240} \, \pi^2 \, \ln^3 2 \, L_t - \frac{2801149}{653184} \, \ln^4 2 \, L_t \\
+ \frac{340853}{680400} \, \ln^5 2 \, L_t - \frac{2801149}{27216} \, \operatorname{Li}_4\left(\frac12\right) \, L_t -\frac{340853}{5670} \, \operatorname{Li}_5\left(\frac12\right) \, L_t \\
+ \frac{85204909}{268800} \, \zeta_3 \, L_t + \frac{33074899}{725760} \, \zeta_5 \, L_t - \frac{4652837}{62208} \, L_t^2 + \frac{2998249}{55296} \, \zeta_3 \, L_t^2 \\
\left. - \frac{32897}{41472} \, L_t^3 - \frac{253}{256} \, L_t^4 \right) \, n_f^2 + \left ( -\frac{1158715}{2985984} -\frac{5}{3888} \, \pi^4 - \frac{8977}{31104} \, \zeta_3 + \frac{10}{27} \, \zeta_ 5 \right. \\
- \frac{185669}{746496} \, L_t +\frac{557341}{7464960} \, \pi^4 \, L_t + \frac{685}{62208} \, \pi^2\ \, \ln^2 2 \, L_t - \frac{685}{62208} \, \ln^4 2 \, L_t \\
- \frac{685}{2592} \, \operatorname{Li}_4\left(\frac12\right) \, L_t - \frac{337009}{55296} \, \zeta_3 \, L_t + \frac{34847}{13824} \, L_t^2 - \frac{28297}{27648} \, \zeta_3 \, L_t^2 - \frac{301}{31104} \, L_t^3 \\
\left. + \frac{245}{3456} \, L_t^4 \right) \, n_f^3 + \left( \frac{65}{31104} -\frac{1}{19440} \, \pi^4 + \frac{5}{1944} \, \zeta_3 + \frac{33349}{279936} \, L_t -\frac{2}{27} \, \zeta_3 \, L_t \right. \\
\left. - \frac{481}{15552} \, L_t^2 + \frac{77}{31104} \, L_t^3 -\frac{1}{648} \, L_t^4 \right) \, n_f^4 - 4 \left(\frac{31}{2} - n_f \right) \, d_{5,0} + 3\left( \beta_6- \beta_6^{\prime} \right) ,
\label{eq:c5}
\end{multline}
where $\zeta_n$ is the Riemann zeta function and $\operatorname{Li}_n(x)$ is the polylogarithm function of order~$n$.

\section{Results from PAs and D-log PAs}\label{app:results}
In this appendix we give the predictions of the Pad\'e and D-log Pad\'e approximants that pass our consistency checks for all the perturbative series analyzed in this work. The approximants marked with $^*$ are the ones that are excluded based on the $\Delta_{n_f}$ criterion given in Eq.~(\ref{eq:Deltanf}). When the exact result is known, it is highlighted in green, while in gray we give the average of the retained approximants with the standard deviation, used as the uncertainty, given in parentheses.

\subsection{Imaginary part of the gluonium correlator}

\begin{table}[H]
\centering
\begin{tabular}{ccccccc}
\thickhline
& $g_{4,1}(1)$ & $g_{4,1}(3)$ & $g_{4,1}(5)$ & $g_{4,1}(7)$ & $g_{4,1}(9)$ & $\Delta_{n_f}$ \\ \hline
$P_2^1(\alpha_s)$ & 102.85 & $-30.121$ & $-87.723$ & $-85.785$ & $-39.827$ & 0.6714 \\
*$P_1^1(u)$ & 251.33 & 47.541 & 10.124 & 767.21 & $-420.08$ & 726.0 \\
$\mathrm{Dlog}_0^2(\alpha_s)$ & 140.97 & 12.271 & $-50.539$ & $-63.334$ & $-40.923$ & 0.0732 \\
*$\mathrm{Dlog}_1^0(u)$ & 110.59 & 4.0936 & 25.557 & 1031.9 & $-612.06$ & 965.7 \\
$\mathrm{Dlog}_0^1(u)$ & 71.86 & $-43.978$ & $-79.257$ & $-27.047$ & 175.15 & 58.68 \\ \midrule
\rowcolor{light-gray}
Average & $105(35)$ & $-21(29)$ & $-73(19)$ & $-59(30)$ & $31(124)$ & --- \\
\midrule
\rowcolor[RGB]{204,255,102}
Exact value & 118.01 & $-15.962$ & $-75.259$ & $-76.998$ & $-37.303$ & --- \\ \thickhline
\end{tabular}
\caption{Estimates of the last known coefficient $g_{4,1}(n_f)$ of the imaginary part of the gluonium correlator, $\operatorname{Im} \Pi_{G^2}$, for different values of $n_f$ from Pad\'e and D-log Pad\'e approximants applied to the perturbative series of $\operatorname{Im} \Pi_{G^2}$ and its Borel transform. The exact value for each $n_f$ is given in green.}
\label{tab:postdictiongluonium}
\end{table}

\begin{table}[H]
\centering
\begin{tabular}{ccccccc}
\thickhline
& $g_{5,1}(1)$ & $g_{5,1}(3)$ & $g_{5,1}(5)$ & $g_{5,1}(7)$ & $g_{5,1}(9)$ & $\Delta_{n_f}$ \\ \hline
$P_2^2(\alpha_s)$ & 72.130 & $-223.79$ & $-197.34$ & 0.082710 & 237.37 & 2.975 \\
*$P_2^1(u)$ & $-326.54$ & $-338.68$ & 6.2620 & 985.16 & 1546.5 & 475.3 \\
*$P_1^2(u)$ & 315.13 & 16.441 & $-1102.2$ & $-323.20$ & $-56.243$ & 12935 \\
$\mathrm{Dlog}_2^1(\alpha_s)$ & 103.87 & $-193.30$ & $-171.34$ & 14.605 & 238.70 & 1.263 \\
$\mathrm{Dlog}_0^3(\alpha_s)$ & 166.44 & $-166.18$ & $-192.74$ & $-57.625$ & 123.09 & 0.1264 \\
*$\mathrm{Dlog}_0^2(u)$ & 326.02 & $-42.232$ & $-12.041$ & 486.20 & 2631.4 & 1771 \\ \midrule
\rowcolor{light-gray}
Average & $114(48)$ & $-194(29)$ & $-187(14)$ & $-14(38)$ & $200(66)$ & --- \\ \thickhline
\end{tabular}
\caption{Estimates of the first unknown coefficient $g_{5,1}(n_f)$ of the imaginary part of the gluonium correlator, $\operatorname{Im} \Pi_{G^2}$, for different values of $n_f$ from Pad\'e and D-log Pad\'e approximants applied to the perturbative series of $\operatorname{Im} \Pi_{G^2}$ and its Borel transform.}
\label{tab:predictiongluonium}
\end{table}

\subsection{Decoupling relation \texorpdfstring{\boldmath{$\zeta_g^2$}}{zeta}}

\begin{table}[H]
\centering
\begin{tabular}{cccccc}
\thickhline
& $d_{4,0}(1)$ & $d_{4,0}(3)$ & $d_{4,0}(5)$ & $d_{4,0}(7)$ & $d_{4,0}(9)$ \\ \hline
$P_1^1(u)$ & 7.7317 & 5.0630 & 2.9571 & 1.4140 & 0.43378 \\
$\mathrm{Dlog}_1^1(a_s)$ & 5.8105 & 3.8089 & 2.2295 & 1.0722 & 0.33701 \\ \midrule
\rowcolor{light-gray}
Average & $6.8(1.4)$ & $4.4(9)$ & $2.6(5)$ & $1.24(24)$ & $0.39(7)$ \\
\midrule
\rowcolor[RGB]{204,255,102}
Exact value & 4.1384 & 1.9428 & $-0.42877$ & $-2.9761$ & $-5.6993$ \\ \thickhline
\end{tabular}
\caption{Estimates of the last known coefficient $d_{4,0}(n_f)$ of the decoupling constant $\zeta_g^2$ for different values of $n_f$ from Pad\'e and D-log Pad\'e approximants applied to the perturbative series of $\zeta_g^2$ and its Borel transform. The exact value for each $n_f$ is given in green.}
\label{tab:postdictiondecoupling}
\end{table}

\begin{table}[H]
\centering
\begin{tabular}{cccccc}
\thickhline
& $d_{5,0}(1)$ & $d_{5,0}(3)$ & $d_{5,0}(5)$ & $d_{5,0}(7)$ & $d_{5,0}(9)$ \\ \hline
$P_2^1(u)$ & 6.3333 & $-11.069$ & $-27.405$ & $-36.595$ & $-32.558$ \\
$P_1^2(u)$ & 25.733 & 7.0079 & 0.44666 & 31.119 & 206.04 \\
$\mathrm{Dlog}_2^1(a_s)$ & 17.246 & 0.47600 & $-13.273$ & $-20.915$ & $-19.362$ \\
$\mathrm{Dlog}_0^3(a_s)$ & 0.13558 & 0.10971 & 0.083844 & 0.057979 & 0.032113 \\
$\mathrm{Dlog}_1^0(u)$ & 18.285 & 3.2561 & 1.6223 & 46.421 & 279.95 \\
$\mathrm{Dlog}_0^1(u)$ & 18.136 & 2.3981 & $-10.162$ & $-17.127$ & $-16.080$ \\ \midrule
\rowcolor{light-gray}
Average & $14(9)$ & $0(6)$ & $-8(11)$ & $0(32)$ & $70(137)$ \\ \thickhline
\end{tabular}
\caption{Estimates of the first unknown coefficient $d_{5,0}(n_f)$ of the decoupling constant $\zeta_g^2$ for different values of $n_f$ from Pad\'e and D-log Pad\'e approximants applied to the perturbative series of $\zeta_g^2$ and its Borel transform.}
\label{tab:predictiondecoupling}
\end{table}

\subsection{QCD \texorpdfstring{\boldmath{$\beta$}}{beta} function}

\begin{table}[H]
\centering
\begin{tabular}{ccccccc}
\thickhline
& $\beta_5(1)$ & $\beta_5(3)$ & $\beta_5(5)$ & $\beta_5(7)$ & $\beta_5(9)$ & $\Delta_{n_f}$ \\ \hline
$\mathrm{Dlog}_1^1(a_s)$ & 1007.3 & 550.10 & 519.14 & $-5.2406$ & 71.828 & 357.6 \\
$\mathrm{Dlog}_0^2(a_s)$ & 369.00 & 160.55 & 45.144 & 14.559 & 74.306 & 12.33 \\
*$\mathrm{Dlog}_1^1(u)$ & 1228.3 & 810.04 & $-3251.1$ & 18.191 & 425.45 & 5019 \\
$\mathrm{Dlog}_0^2(u)$ & 943.08 & 448.19 & 165.45 & 92.515 & 493.75 & 216.6 \\ \midrule
\rowcolor{light-gray}
Average & $773(351)$ & $386(202)$ & $243(246)$ & $34(52)$ & $213(243)$ & --- \\
\midrule
\rowcolor[RGB]{204,255,102}
Exact value & 719.38 & 254.64 & 30.216 & 21.645 & 203.10 & --- \\ \thickhline
\end{tabular}
\caption{Estimates of the last known coefficient $\beta_5(n_f)$ of the $\beta$ function, for different values of $n_f$ from Pad\'e and D-log Pad\'e approximants applied to the perturbative series of the $\beta$ function and its Borel transform. The exact value for each $n_f$ is given in green.}
\label{tab:postdictionbeta}
\end{table}

\begin{table}[H]
\centering
\begin{tabular}{ccccccc}
\thickhline
& $\beta_6(1)$ & $\beta_6(3)$ & $\beta_6(5)$ & $\beta_6(7)$ & $\beta_6(9)$ & $\Delta_{n_f}$ \\ \hline
$P_2^3(a_s)$ & 3168.9 & 1121.9 & 300.33 & $-90.148$ & 67.152 & 11.26 \\
$\mathrm{Dlog}_2^1(a_s)$ & 3948.8 & 1660.0 & 500.50 & $-101.08$ & 90.724 & 5.162 \\
*$\mathrm{Dlog}_1^2(a_s)$ & 2857.6 & 654.35 & 39.312 & $-13.833$ & $-2792.1$ & 4344 \\
$\mathrm{Dlog}_0^3(a_s)$ & 1736.0 & 513.63 & 32.139 & $-18.419$ & $-72.067$ & 6.921 \\
*$\mathrm{Dlog}_1^2(u)$ & 4660.5 & 1907.6 & 388.63 & $-163.54$ & $-1995.8$ & 7831 \\
*$\mathrm{Dlog}_0^3(u)$ & 3261.2 & 583.38 & $-380.91$ & $-496.06$ & 1768.1 & 1099 \\ \midrule
\rowcolor{light-gray}
Average & $2951(1100)$ & $1099(574)$ & $278(235)$ & $-70(45)$ & $29(88)$ & --- \\ \thickhline
\end{tabular}
\caption{Estimates of the first unknown coefficient $\beta_6(n_f)$ of the $\beta$ function, for different values of $n_f$ from Pad\'e and D-log Pad\'e approximants applied to the perturbative series of the $\beta$ function and its Borel transform.}
\label{tab:predictionbeta}
\end{table}

\subsection{Imaginary part of the scalar correlator}

\begin{table}[H]
\centering
\begin{tabular}{ccccccc}
\thickhline
& $p_4(1)$ & $p_4(3)$ & $p_4(5)$ & $p_4(7)$ & $p_4(9)$ & $\Delta_{n_f}$ \\ \hline
$P_1^2(\alpha_s)$ & $-2.2697$ & $-6.4318$ & $-8.7955$ & $-9.3020$ & $-7.8079$ & 11.07 \\
$P_1^1(u)$ & 2.3139 & $-4.2649$ & $-8.1645$ & $-9.2918$ & $-7.4267$ & 16.58 \\
$\mathrm{Dlog}_2^0(\alpha_s)$ & $-2.7654$ & $-6.3290$ & $-8.9025$ & $-10.568$ & $-11.408$ & 7.144 \\
$\mathrm{Dlog}_1^1(\alpha_s)$ & 0.36633 & $-5.6071$ & $-8.8669$ & $-8.5333$ & $-2.1116$ & 169.9 \\
$\mathrm{Dlog}_0^2(\alpha_s)$ & $-7.4426$ & $-8.3145$ & $-9.0257$ & $-9.5859$ & $-10.005$ & 0.0374 \\
$\mathrm{Dlog}_1^0(u)$ & 1.7679 & $-5.1694$ & $-8.9567$ & $-9.4346$ & $-6.2746$ & 25.27 \\
$\mathrm{Dlog}_0^1(u)$ & $-1.7886$ & $-5.9048$ & $-8.9979$ & $-11.150$ & $-12.443$ & 7.144 \\ \midrule
\rowcolor{light-gray}
Average & $-1.4(3.3)$ & $-6.0(1.3)$ & $-8.82(30)$ & $-9.7(9)$ & $-8.2(3.5)$ & --- \\
\midrule
\rowcolor[RGB]{204,255,102}
Exact value & $-1.7650$ & $-5.5112$ & $-8.4771$ & $-10.673$ & $-12.108$ & --- \\ \thickhline
\end{tabular}
\caption{Estimates of the last known coefficient $p_4(n_f)$ of $\operatorname{Im} \Pi(s)$ for different values of $n_f$ from Pad\'e and D-log Pad\'e approximants applied to the perturbative series of the second derivative of the scalar correlator, $\Pi^{''}$, and its Borel transform. The exact value for each $n_f$ is given in green.}
\label{tab:postdictionhbb}
\end{table}

\begin{table}[H]
\centering
\begin{tabular}{ccccccc}
\thickhline
& $p_5(1)$ & $p_5(3)$ & $p_5(5)$ & $p_5(7)$ & $p_5(9)$ & $\Delta_{n_f}$ \\ \hline
$P_2^2(\alpha_s)$ & $-37.516$ & $-31.849$ & $-27.955$ & $-20.052$ & $-12.478$ & 1107 \\
*$P_2^1(u)$ & $-84.230$ & $-31.806$ & $-26.788$ & $-19.715$ & $-5.5800$ & 19464 \\
$\mathrm{Dlog}^0_3(\alpha_s)$ & $-36.693$ & $-32.781$ & $-27.021$ & $-19.393$ & $-10.002$ & 45.88 \\
$\mathrm{Dlog}^2_1(\alpha_s)$ & $-37.142$ & $-31.764$ & $-26.693$ & $-21.455$ & $-15.741$ & 61.23 \\
$\mathrm{Dlog}^3_0(\alpha_s)$ & $-50.943$ & $-37.980$ & $-27.781$ & $-19.571$ & $-12.682$ & 39.78 \\
$\mathrm{Dlog}^0_2(u)$ & $-43.900$ & $-31.079$ & $-26.628$ & $-16.156$ & 21.660 & 2596 \\
$\mathrm{Dlog}^2_0(u)$ & $-32.716$ & $-31.157$ & $-26.250$ & $-18.120$ & $-7.0220$ & 48.67 \\ \midrule
\rowcolor{light-gray}
Average & $-40(7)$ & $-33(3)$ & $-27.1(7)$ & $-19(2)$ & $-6(14)$ & --- \\ \thickhline
\end{tabular}
\caption{Estimates of the first unknown coefficient $p_5(n_f)$ of the imaginary part of the scalar correlator, $\operatorname{Im} \Pi(s)$, for different values of $n_f$ from Pad\'e and D-log Pad\'e approximants applied to the perturbative series of the second derivative of the scalar correlator, $\Pi^{''}$, and its Borel transform.}
\label{tab:predictionhbb}
\end{table}

\section{Fit results with more digits}\label{app:fits}
In this Appendix, we report the covariance matrix and the fit parameters obtained for the first unknown coefficient of the quantities analyzed in this work, i.e. the imaginary part of the gluonium correlator, $\operatorname{Im} \Pi_{G^2}(s)$, the decoupling constant $\zeta_g^2$, the QCD $\beta$ function, and the imaginary part of the scalar correlator, $\operatorname{Im} \Pi(s)$, with additional digits for the reproduction of the derived coefficients.

\subsection{Imaginary part of the gluonium correlator}
The parameters of the fit, which are the coefficients $g_{5,1}^{(k)}$, are estimated as
\begin{align}
g_{5,1}^{(0)} = 441.257808, &\qquad \qquad g_{5,1}^{(1)} = -396.647125, \quad \quad g_{5,1}^{(2)} = 73.7546786, \nonumber \\
g_{5,1}^{(3)} &= -4.27837426, \qquad \qquad g_{5,1}^{(4)} = 0.0746385362.
\end{align}
The covariance matrix of these parameters is given in Tab.~\ref{tab:covarianceg5}.
\begin{table}[h]
\small
\setlength{\tabcolsep}{4pt}
\centering
\begin{tabular}{lrrrrr}
\toprule
& $g_{5,1}^{(0)}$ & $g_{5,1}^{(1)}$ & $g_{5,1}^{(2)}$ & $g_{5,1}^{(3)}$ & $g_{5,1}^{(4)}$ \\ \midrule
$g_{5,1}^{(0)}$ & 2943.79175 & & & & \\
$g_{5,1}^{(1)}$ & $-276.623858$ & 26.5180258 & & & \\
$g_{5,1}^{(2)}$ & $-62.3184661$ & 4.73810163 & 3.70369305 & & \\
$g_{5,1}^{(3)}$ & $-2.54356064$ & 0.394214286 & $-0.277197464$ & 0.0481579707 & \\
$g_{5,1}^{(4)}$ & 0.464464150 & $-0.0502139190$ & 0.00417910689 & $-0.00234660464$ & 0.000155617217 \\ \bottomrule
\end{tabular}
\caption{Estimated covariance matrix of the fit parameters $g_{5,1}^{(k)}$ governing the $n_f$-dependence of the fifth-order coefficient of $\operatorname{Im} \Pi_{G^2}$. Only the lower triangular portion is reported.}
\label{tab:covarianceg5}
\end{table}

\subsection{Decoupling relation \texorpdfstring{\boldmath{$\zeta_g^2$}}{zeta}}
The parameters of the fit, which are the coefficients $d_{5,0}^{(k)}$, are estimated as
\begin{align}
d_{5,0}^{(0)} &= 17.6331298, \qquad \qquad \,\,\,\, d_{5,0}^{(1)} = -1.79262194, \nonumber \\
d_{5,0}^{(2)} &= -2.27375222, \qquad \qquad d_{5,0}^{(3)} = 0.318431039.
\end{align}
The covariance matrix of these parameters is given in Tab.~\ref{tab:covarianced50}.
\begin{table}[h]
\centering
\begin{tabular}{lrrrr}
\toprule
& $d_{5,0}^{(0)}$ & $d_{5,0}^{(1)}$ & $d_{5,0}^{(2)}$ & $d_{5,0}^{(3)}$ \\ \midrule
$d_{5,0}^{(0)}$ & 112.647973 & & & \\
$d_{5,0}^{(1)}$ & $-26.7304602$ & 52.1839583 & & \\
$d_{5,0}^{(2)}$ & $-2.67615730$ & $-19.8240819$ & 9.73383082 & \\
$d_{5,0}^{(3)}$ & 0.555983330 & 2.54876889 & $-1.27261451$ & 0.166865675 \\ \bottomrule
\end{tabular}
\caption{Estimated covariance matrix of the fit parameters $d_{5,0}^{(k)}$ governing the $n_f$-dependence of the fifth-order coefficient of $\zeta_g^2$. Only the lower triangular portion is reported.}
\label{tab:covarianced50}
\end{table}

\subsection{QCD \texorpdfstring{\boldmath{$\beta$}}{beta} function}
The parameters of the fit, which are the coefficients $\beta_6^{(k)}$, are estimated as
\begin{align}
\beta_6^{(0)} = 3756.92176, &\qquad \qquad \beta_6^{(1)} = -1024.80867, \quad \quad \beta_6^{(2)} = 38.3490168, \nonumber \\
\beta_6^{(3)} &= 8.06786223, \qquad \qquad \beta_6^{(4)} = -0.546105755.
\end{align}
The covariance matrix of these parameters is given in Tab.~\ref{tab:covariancebeta6}.
\begin{table}[h]
\small
\centering
\begin{tabular}{lrrrrr}
\toprule
& $\beta_6^{(0)}$ & $\beta_6^{(1)}$ & $\beta_6^{(2)}$ & $\beta_6^{(3)}$ & $\beta_6^{(4)}$ \\ \midrule
$\beta_6^{(0)}$ & 925231.691 & & & & \\
$\beta_6^{(1)}$ & 75280.6392 & 44634.1245 & & & \\
$\beta_6^{(2)}$ & $-85110.8026$ & $-21489.5513$ & 13337.7393 & & \\
$\beta_6^{(3)}$ & 9440.78633 & 2618.00890 & $-1568.08865$ & 185.193620 & \\
$\beta_6^{(4)}$ & $-228.233177$ & $-97.3738711$ & 50.7994917 & $-6.11434924$ & 0.217562258 \\ \bottomrule
\end{tabular}
\caption{Estimated covariance matrix of the fit parameters $\beta_6^{(k)}$ governing the $n_f$-dependence of the sixth-order coefficient of the $\beta$ function. Only the lower triangular portion is reported.}
\label{tab:covariancebeta6}
\end{table}

\subsection{Imaginary part of the scalar correlator}
The parameters of the fit, which are the coefficients $p_5^{(k)}$, are estimated as
\begin{align}
p_5^{(0)} &= -44.5007207, \qquad \qquad \,\,\, p_5^{(1)} = 5.46275357, \nonumber \\
p_5^{(2)} &= -0.714679511, \qquad \qquad p_5^{(3)} = 0.0647831764.
\end{align}
The covariance matrix of these parameters is given in Tab.~\ref{tab:covariancehbb}.
\begin{table}[h]
\centering
\begin{tabular}{lrrrr}
\toprule
& $p_5^{(0)}$ & $p_5^{(1)}$ & $p_5^{(2)}$ & $p_5^{(3)}$ \\ \midrule
$p_5^{(0)}$ & 149.213221 & & & \\
$p_5^{(1)}$ & $-87.2887898$ & 62.7170491 & & \\
$p_5^{(2)}$ & 17.5902255 & $-14.3038614$ & 3.46044164 & \\
$p_5^{(3)}$ & $-1.19632079$ & 1.04667206 & $-0.261305064$ & 0.0200706707 \\\bottomrule
\end{tabular}
\caption{Estimated covariance matrix of the fit parameters $p_5^{(k)}$ governing the $n_f$-dependence of the fifth-order coefficient of $\operatorname{Im} \Pi(s)$. Only the lower triangular portion is reported.}
\label{tab:covariancehbb}
\end{table}

\addcontentsline{toc}{section}{References}
\bibliographystyle{jhep}
\bibliography{References}

\end{document}